\documentclass[aps,prl,twocolumn,longbibliography,superscriptaddress]{revtex4-2}

\usepackage[usenames,dvipsnames]{color}

\usepackage{amsfonts}
\usepackage{amsmath}
\usepackage{comment}
\usepackage{cancel}
\usepackage{ulem}
\usepackage{physics}
\usepackage{graphicx}
\usepackage{dcolumn}
\usepackage{bm}
\usepackage{gensymb}
\usepackage{tensor}
\usepackage{rotating}
\usepackage[percent]{overpic}
\usepackage{array}
\usepackage{multibib}
\usepackage{comment}
\usepackage{blkarray}       

\def\be{\begin{equation}}
\def\ee{\end{equation}}
\def\ber{\begin{eqnarray}}
\def\eer{\end{eqnarray}}

\usepackage[breaklinks=true,colorlinks,citecolor=blue,linkcolor=blue,urlcolor=blue]{hyperref}

\begin{document}

\title{\textbf{Local-moment magnets on all close-packed lattices are equivalent} 
}

\author{Sam Studdy}
\email{samstuddy3@gmail.com}
\affiliation{Department of Physics, Brock University, St. Catharines, Ontario L2S 3A1, Canada}
\author{Alwyn Jose Raja}
\email{alwynjoseraja2000@gmail.com}
\affiliation{Department of Physics, Indian Institute of Technology Madras, Chennai 600036, India}
\author{R. Ganesh}
\email{r.ganesh@brocku.ca}
\affiliation{Department of Physics, Brock University, St. Catharines, Ontario L2S 3A1, Canada}

\date{\today}

\begin{abstract}
The Heisenberg antiferromagnet on the fcc lattice is a canonical example of frustration. The fcc lattice is but one of an infinite number of close-packed structures, all of which have the same frustration-inducing local geometry. In this letter, we demonstrate a hidden symmetry that relates classical magnets on all close-packed lattices. This symmetry derives from a similar property in tight binding models that may 
explain the disordered low-temperature structures of lithium and sodium metals. In magnets, the symmetry holds when couplings are restricted to nearest ($J_1$) and next-nearest neighbours ($J_2$). It guarantees that \textit{all} close-packed lattices are equivalent at the Luttinger-Tisza level.
For any given values of $J_1$ and $J_2$, they have the same ground-state energy. Their ground state momenta are identical when projected onto the stacking plane. We construct a unified phase diagram as a function of $J_1$ and $J_2$ couplings for Heisenberg, XY and Ising magnets on any close-packed lattice. In each phase, the ground-state degeneracy is 
the same on all close-packed lattices, even after accounting for multi-$Q$ spirals.  This phase diagram can be easily adapted to any magnet composed of stacked triangular layers with similar couplings. 
Fluctuations above the classical ground state(s) are not related by any symmetry. This suggests that close-packed structures can show interesting differences in order-by-disorder selection. 
\end{abstract}

\maketitle

\noindent{\color{blue} \it{Introduction}.}---  
The fcc lattice can be viewed as a network of corner-sharing tetrahedra. With antiferromagnetic couplings, this geometry ensures that all bonds cannot be satisfied simultaneously. In 1950, Anderson demonstrated that the classical fcc Heisenberg antiferromagnet has infinite ground state degeneracy\cite{Anderson1950}. Since then, a large body of work has emerged on its ordering properties\cite{Luttinger1951,Ziman1953,Lines1962,Lines1964,Yamamoto1972,Yamamoto1972,Oguchi1985,Henley1987,Zhitomirsky2020,Zhitomirsky2022,Oitmaa2023}. Several papers have studied this model with next-nearest and third-neighbour couplings, finding new phases and degeneracies\cite{Oja1993,Sun2018,Balla2020}. The lasting interest in fcc magnets also draws from several material realizations, including NiO, MnO\cite{shull1951neutron} and double perovskites\cite{Cava2011}. 
Fcc is not unique in its networked-tetrahedral-structure. 
The hcp lattice is similar  and also shows frustration\cite{Aplesnin1989,Diep1992,Diep2012,Niggemann2020}. Fcc and hcp are the simplest members of the infinite set of close-packed lattices, all of which can be constructed by linking tetrahedra. In this letter, we demonstrate an equivalence across classical magnets on all close-packed lattices.

We build on recent work by one of us in the context of lithium and sodium metals\cite{He2024}. These two elemental solids are anomalous among crystalline materials, as they exhibit an order-to-disorder transition upon cooling\cite{barrett_PR_1947,Barrett1955}. At low-temperatures, they are `structurally frustrated' as all close-packed structures are nearly iso-energetic\cite{ackland_Science_2017}. This stems from a gauge symmetry that connects their Bloch Hamiltonians\cite{He2024}. Here, we use the same construction in classical magnets with a symmetry that connects Luttinger Tisza (LT) matrices. 

\noindent{\color{blue} \it{Geometry of close-packing}.}---
\label{sec.closepacking}
Close-packed structures\cite{Szpiro2003kepler,KrishnaPandey1981} are solutions to the sphere-packing problem. They are obtained by stacking triangular-lattice layers. Each structure can be represented as a Barlow sequence: a string of letters  A, B or C, with the constraint that adjacent letters cannot be the same. Alternatively, it can be represented as a H\"agg code\cite{Hagg1943} -- a 1D Ising configuration, where each entry represents the shift from one layer to the next. A forward shift ($A\rightarrow B$, $B\rightarrow C$ or $C\rightarrow A$) is denoted as `$+$', while a backward shift ($A\rightarrow C$, $B\rightarrow A$ or $C\rightarrow B$) is represented as `$-$'. For example, the fcc lattice is represented as (ABC) or $(+)$, whereas the hcp lattice is expressed as (AB) or $(+-)$. The parentheses indicate that the pattern within repeats indefinitely. 

\begin{figure}
    \centering
    \includegraphics[width=0.85\columnwidth]{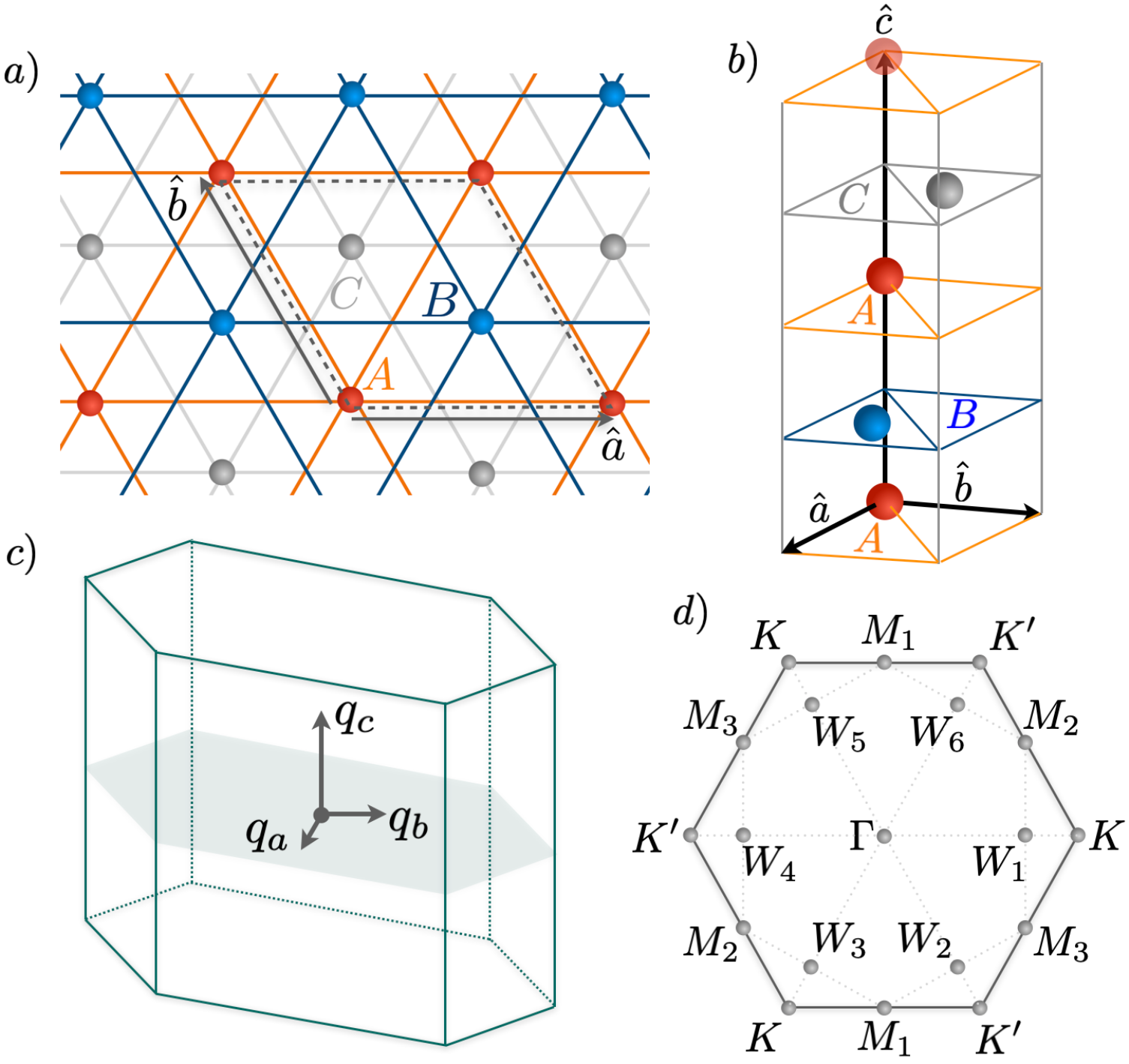}
    \caption{a) A close-packed structure viewed along the stacking axis. Generically, the unit cell forms a rhombus, with primitive translations $\hat{a}$ and $\hat{b}$ as shown. b) As an example, the unit cell of the ABAC structure is shown. The primitive translation vector $\hat{c}$ is parallel to the stacking axis. c) The Brillouin zone of an $M$-layer structure forms a hexagonal prism or (d) a hexagon when projected onto the stacking plane. Momenta of interest are labeled. 
   }
    \label{fig.unitcell}
\end{figure}

In every close-packed lattice, each site has twelve nearest neighbours -- six within the plane, three above and three below. There are six next-nearest neighbours -- three each in the planes above and below. Third neighbours vary from one stacking arrangement to another\cite{He2024}. All close-packed structures have translational symmetry perpendicular to the stacking axis, with the translation vectors of a triangular lattice. Below, we consider a generic close-packed structure with a repeating sequence of $M$ layers. This gives rise to an additional primitive translation vector, $\hat{c}$, along the stacking axis. Any stacking sequence can be described using the unit cell shown in Fig.~\ref{fig.unitcell}(a,b), giving rise to the hexagonal prism of Fig.~\ref{fig.unitcell}(c) as the Brillouin zone. In all close-packed structures, the projection of the Brillouin zone onto the stacking plane yields a hexagon, as shown in Fig.~\ref{fig.unitcell}(d). 

\noindent{\color{blue} \it{Model and Luttinger-Tisza analysis}.}---
We consider a Heisenberg model with spins at the sites of a generic close-packed lattice. We later argue that our results also apply to XY and Ising magnets. We restrict couplings to nearest-neighbours $(J_1)$ and next-nearest neighbours $(J_2)$, with arbitrary values for $J_1$ and $J_2$. In the classical limit, the Hamiltonian is given by
\begin{eqnarray}
    H = J_1 \sum_{\langle ij \rangle} \mathbf{S}_i \cdot \mathbf{S}_j + J_2 \sum_{\langle\langle ij \rangle\rangle} \mathbf{S}_i \cdot \mathbf{S}_j.
\end{eqnarray}
In the celebrated Luttinger-Tisza (LT) approach\cite{luttinger1946theory}, ground states are found by optimizing over Fourier-transformed spin components. The constraint of fixed-spin-length is imposed on average. An LT matrix is constructed at each momentum and its eigenvalues are determined.
The smallest eigenvalue over all momenta yields a lower bound for the ground state energy. If the corresponding eigenvector satisfies a `pure-phases' condition, it yields the true classical ground state energy. That is, 
the bound can be saturated by a legitimate spin arrangement. This condition stipulates that all entries must have the same magnitude, while the phases may differ. A detailed description is provided in the supplementary material, with an explicit example\cite{supplementary}. 

For periodic stacking with an $M$-layer unit cell, we first express the stacking sequence in H\"agg code notation as $(\sigma_1,\sigma_2,\ldots,\sigma_M)$, where the $\sigma$'s are Ising variables. The LT matrix takes a tridiagonal form,
\begin{eqnarray}
M_{\{\sigma\},\mathbf{q}_{\parallel},q_c}
=
    \left(
\begin{array}{cccccc}
V_D & V_{1} & 0 & 0 & \ldots & V_{M}^* \\
V_{1}^* & V_{D} & V_2 & 0 & \ddots &0 \\
0 & V_{2}^* & V_{D} & V_3 & \ddots & 0  \\
 0 & 0 & \ddots & \ddots & \ddots & 0 \\
 0 & 0 & \ddots & \ddots & V_{D} & V_{M-1} \\
V_{M} & 0 & \ldots & 0 & V_{M-1}^* & V_{D}
\end{array}
    \right).
\label{eq.Mmatrix}
\end{eqnarray}

A detailed derivation is provided in the supplementary material\cite{supplementary}. The matrix is written in the layer basis, i.e., the $(i,j)^\mathrm{th}$ element represents the coupling between spins in layer $i$ and those in layer $j$. The entries are functions of momentum coordinates: $q_c$, momentum along the stacking axis, and $\mathbf{q}_\parallel$, the projection onto the stacking plane. All diagonal entries are identical,  
\begin{eqnarray}
    V_D (\mathbf{q}_\parallel) =  J_1 \{ \cos q_a +\cos q_b + \cos (q_a+q_b) \}  - \lambda,
    \label{eq.Vdiagonal}
\end{eqnarray}
encoding couplings within each plane. Here, $\lambda$ is a Lagrange multiplier that enforces the spin-length constraint. The right-of-diagonal entry $V_j$ takes one of two values, depending on the H\"agg code entry $\sigma_j$,
\begin{eqnarray}
 V_{j}(\sigma_j,\mathbf{q}_\parallel,q_c) = \frac{e^{i {q}_c/M}}{2} \bar{f}(\mathbf{q}_\parallel) \exp\big\{ i \sigma_j  \phi(\mathbf{q}_\parallel)  \big\},
\label{eq.Vj}
\end{eqnarray}
 
Here, $\bar{f}$ and $\phi$ are the amplitude and phase of the function
\begin{eqnarray}
f(\mathbf{q}_\parallel) = J_1 \sum_{\mathbf{u}} e^{i\mathbf{q}_\parallel \cdot \mathbf{u} } +J_2 \sum_{\mathbf{v}} e^{i\mathbf{q}_\parallel \cdot \mathbf{v} } \equiv \bar{f}(\mathbf{q}_\parallel) e^{ i\phi(\mathbf{q}_\parallel) }, 
\label{eq.phidef}
\end{eqnarray}
where $\mathbf{u} \in \{\frac{2\hat{a}}{3}+\frac{\hat{b}}{3}, -\frac{\hat{a}}{3}-\frac{2\hat{b}}{3},-\frac{\hat{a}}{3}+\frac{\hat{b}}{3}\}$, in-plane displacements to reach nearest neighbours of the next layer when $\sigma_j = +1$. Similarly, $\mathbf{v} \in \{\frac{2\hat{a}}{3}+\frac{4\hat{b}}{3}, \frac{2\hat{a}}{3}-\frac{2\hat{b}}{3},-\frac{4\hat{a}}{3}-\frac{2\hat{b}}{3}\}$, corresponding to next-nearest-neighbour displacements. 

\noindent{\color{blue} \it{Gauge symmetry across close-packed structures}.}---We consider LT matrices for two distinct close-packed structures with repeating $M$-layer stacking sequences. We represent them  in H\"agg code notation as $(\sigma_1,\ldots,\sigma_M)$ and $(\chi_1,\ldots,\chi_M)$. Remarkably, the two LT matrices are related by a unitary transformation,
\begin{eqnarray}
   \mathcal{U}^\dagger~ M_{\{\sigma\},\mathbf{q}_{\parallel},q_c}~ \mathcal{U}=  
    M_{\{\chi\},\mathbf{q}_{\parallel},q_c+Q},
    \label{eq.unitary}
\end{eqnarray}
where $\mathcal{U}=\mathrm{Diag}\{e^{i\theta_1},e^{i\theta_2},\ldots,e^{i\theta_M}\} $, with 
\begin{eqnarray}
Q &=& \phi(\mathbf{q}_\parallel) 
\sum_{j=1}^M (\sigma_j - \chi_j); 
\label{eq.Qdef}\\
    \theta_k &=& (k-1)\frac{Q}{M} + \phi(\mathbf{q}_\parallel) 
\sum_{j=1}^{k-1} (\sigma_j - \chi_j),~~k=1,\ldots,M.~~~~~~
\end{eqnarray}
Here, $\phi(\mathbf{q}_\parallel)$ is the phase defined in Eq.~\ref{eq.phidef}. In the supplementary material\cite{supplementary}, we provide an explicit example and a rationalization in terms of the Aharanov-Bohm effect in layer space. 

The unitary relation in Eq.~\ref{eq.unitary} has two consequences:

(i) The LT eigenvalues of both structures are identical. In particular, they are identical at each $\mathbf{q}_\parallel$. The momentum along the stacking axis, $q_c$, must be shifted by $Q$ to have the eigenvalues matched. This shift is not required in certain cases. For instance, $Q$ vanishes if 
the two structures have the same `net moment', i.e., if $\sum \sigma_j = \sum \chi_j$. 

(ii) The unitary matrix $\mathcal{U}$ consists solely of pure phases along the diagonal. This ensures that a pure-phases eigenvector of one structure is related to a pure-phases eigenvector of the second. That is, if the LT minimum of one structure yields a legitimate ground state, so does the LT minimum of the other. 

The equivalence relation in Eq.~\ref{eq.unitary} connects two structures with the same number of layers. It can be easily generalized to relate a structure with $M_1$ layers to another with $M_2$ layers. This can be achieved by considering an expanded unit cell that accommodates both structures --- with $\mathrm{lcm}(M_1,M_2)$ layers where $\mathrm{lcm}$ denotes the least common multiple. 

We conclude that the $J_1-J_2$ Heisenberg model on \textit{any} close-packed lattice has
(i) the same minimum energy from LT analysis, and (ii) the same LT minimum (or minima) in momentum space when projected onto the plane perpendicular to the stacking direction, i.e., as a function of $\mathbf{q}_\parallel$. Below, we explicitly construct legitimate spin configurations to establish that the LT minimum is indeed the true classical ground state energy.

\begin{figure}[htbp]
    
    \begin{minipage}[c]{0.5\columnwidth}
        \centering
        \includegraphics[width=\linewidth]{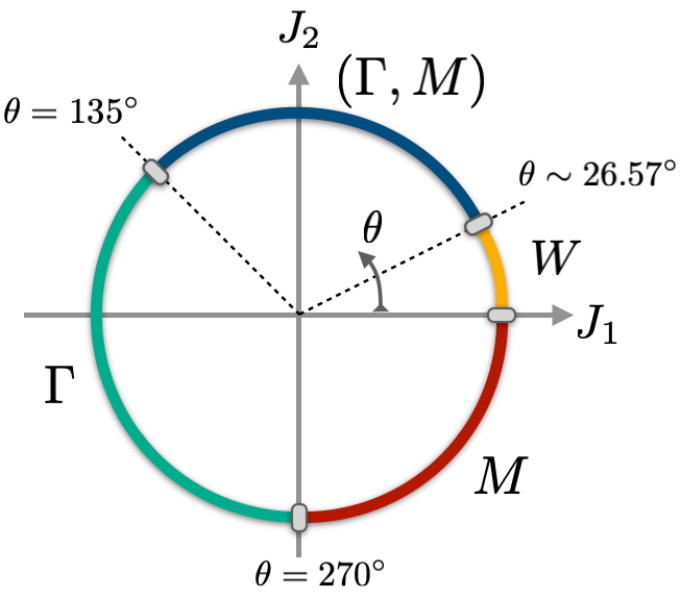}
    \end{minipage}
    \hfill 
    \begin{minipage}[c]{0.45\columnwidth}
        \centering
        \begin{tabular}{|c|c|}
         \hline Phase & E$_{g.s.}$ \\ \hline\hline
         $\Gamma$ &  $6J_1 + 3J_2$  \\ \hline
         $M$ & $-2J_1 + 3J_2$  \\ \hline
         $W$ & $-2J_1 + J_2$ \\ \hline
         $(\Gamma,M)$ & $-3J_2$ \\ \hline
    \end{tabular}
    \end{minipage}
    
    \caption{Unified phase diagram for the $J_1-J_2$ model on any close-packed lattice. Phases are labeled by the location(s) of LT minimum (or minima) within the projected Brillouin zone. The table shows the classical ground state energy, $E_{g.s.
    }$, on a per-spin basis with the spin length $S$ set to unity.}
    \label{fig.phasediagram}
\end{figure}

\noindent{\color{blue} \it{Unified phase diagram for all close-packed lattices}.}--- 
Upon varying $J_1$ and $J_2$ values, LT analysis yields four phases. They are shown in Fig.~\ref{fig.phasediagram}, with the two coupling constants parameterized as ($J_1, J_2$)$\equiv(\cos\theta,\sin\theta)$, up to an overall energy scale. We label the phases as $\Gamma$, $M$, $W$ and $(\Gamma,M)$ to reflect the locations of LT minima in the projected Brillouin zone. 

\begin{figure*}
\includegraphics[width=\linewidth]{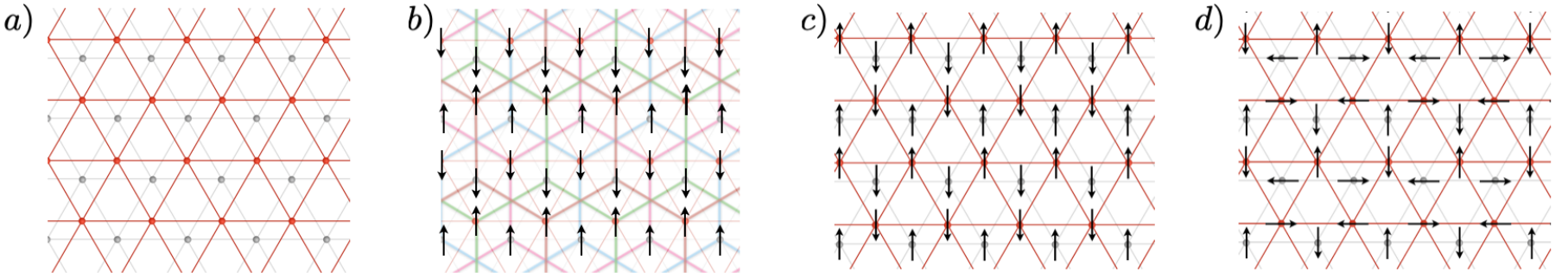}
    \caption{a) Two adjacent layers in a close-packed structure. b) Four sublattices spanned by $J_2$ bonds. The four-sublattice structure persists for any number of layers. c) A single-$Q$ ground state in the $M$ phase. Spin orientations are shown in two layers. In subsequent layers, spins are fixed such that each $J_2$-sublattice is ferromagnetically ordered. A single-$Q$ ground state in the $(\Gamma,M)$ phase is similar, with each $J_2$-sublattice ordered \textit{antiferromagnetically}. d) A single-$Q$ ground state in the $W$ phase. Spins in other layers are oriented to satisfy the tetrahedron rule -- see text.}
    \label{fig.singleQ}
\end{figure*}

Crucially, all four phases support legitimate single-$Q$ ground state configurations on any close packed lattice. We illustrate this in Fig.~\ref{fig.singleQ} on two adjacent layers. The structures on further layers can be obtained by an appropriate rule for each phase, as we describe next. It can be easily seen that these rules yield the correct classical energy for any stacking sequence.

The $\Gamma$ phase is particularly simple, as all spins are parallel. To describe orientations in the $M$ and $(\Gamma,M)$ phases, we first discuss the connectivity of the next-nearest neighbour bonds. In any close-packed lattice, the $J_2$ bonds form four interpenetrating sublattices as shown in Fig.~\ref{fig.singleQ}(b). Each $J_2$ sublattice is bipartite. This can be seen by  
dividing the system into even- and odd-numbered layers. Every $J_2$ bond connects a site from an even layer to an odd layer. In the $M$ phase, the spin configuration on layers is such that each $J_2$ sublattice is ferromagnetically ordered. In the $(\Gamma,M)$ phase, each sublattice orders antiferromagnetically. 

In the $W$ phase, spins on all layers orient so as to satisfy a `tetrahedron rule'. On any close-packed lattice, we can form tetrahedra by combining a triangle on one layer (three neighbouring sites) with an apex site that sits above (or below) the centroid in an adjacent layer. In the $W$ phase, each tetrahedron has four spins that add to zero. This rule has previously been discussed on the fcc lattice\cite{Balla2020}. The same operates on all close packings.

\noindent{\color{blue} \it{Multi-$Q$ states.}}--- 
Remarkably, three of the four phases allow for multi-$Q$ orders. A generic ground state can be constructed by combining LT minima in momentum space, as shown in the Supplementary Material\cite{supplementary}. Here, we work in real space and construct generic ground states using suitable motifs. Crucially, the resulting ground state degeneracy is the same on all close-packed lattices. For example, in the $(\Gamma,M)$ phase, a generic ground state is formed by the four $J_2$-sublattices forming independent N\'eel  antiferromagnets. The set of classical ground states is parameterized by four N\'eel vectors -- regardless of stacking sequence. 

\begin{figure}
    \centering
    \includegraphics[width=\linewidth]{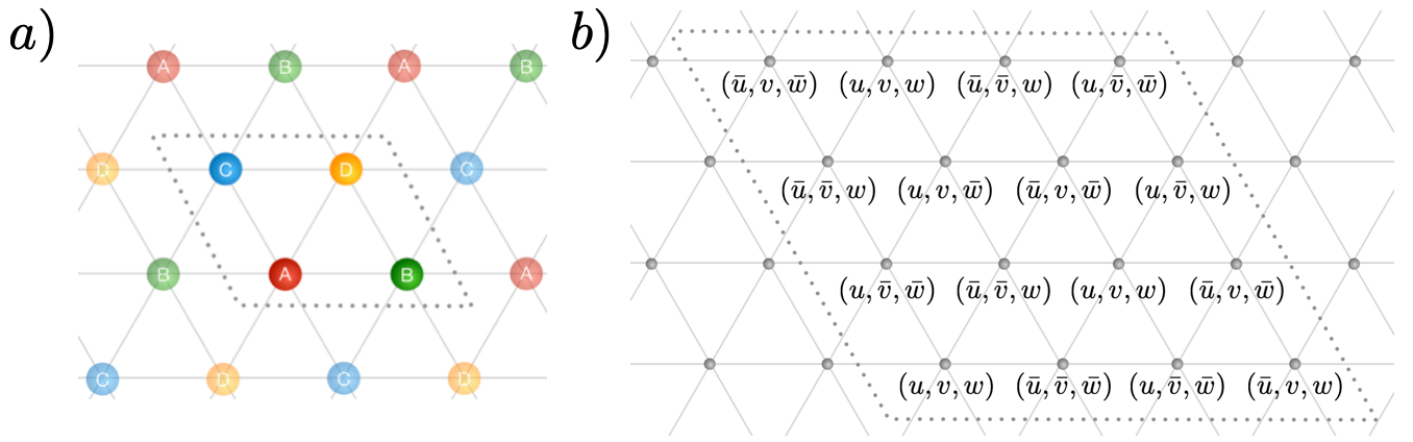}
    \caption{Multi-$Q$ ground states. a) A generic ground state in the $X$ phase with a four-site motif in one layer. Spins in other layers are fixed such that each $J_2$ sublattice is ferromagnetically ordered. b) A generic ground state in the $W$ phase, with a sixteen-site motif within a layer. Spins in other layers are fixed by the tetrahedron rule.} 
    \label{fig.multiQ}
\end{figure}

In the $M$ phase, the generic ground state carries a four-site motif on any one layer as shown in  Fig.~\ref{fig.multiQ}(a). On the four sites labeled as $A$, $B$, $C$ and $D$, we choose four spin directions subject to the condition that $\mathbf{S}_A + \mathbf{S}_B +\mathbf{S}_C + \mathbf{S}_D=0 $. 

Spins on other layers are obtained from the condition that each $J_2$ sublattice forms a ferromagnet. The degeneracy amounts to the number of ways of choosing four unit vectors such that their sum is zero. This is a well-known problem in several contexts within frustrated magnetism\cite{Litvin1974,Reimers1992,Canals1998,Moessner_Chalker1998,Henley2005,Henley2010,Pandey2020}, resulting in a five-dimensional space with singularities at collinear states\cite{Khatua_quadrumer}.

In the $W$ phase, a generic ground state is described by a sixteen-site motif, shown in Fig.~\ref{fig.multiQ}(b). 
Spin components are described by three real numbers $u$, $v$ and $w$ which satisfy the normalization condition $u^2+v^2+w^2=1$. The figure shows the spin configuration within a single layer. Spins on the adjacent layer are chosen to satisfy the tetrahedron rule. This rule can be iteratively used to fix spin orientations on all layers. The space of ground states is five-dimensional -- parametrized by three Euler angles (to choose a coordinate frame) and three real numbers ($u$, $v$ and $w$) that are subject to one normalization constraint.

\noindent{\color{blue} \it{Phase boundaries in the $J_1-J_2$ model}.}---
The phase boundaries in Fig.~\ref{fig.phasediagram} show interesting LT manifolds and ground state degeneracies. The boundary at $\theta=270^\circ$ is the simplest, with LT minima at $\Gamma$ and $M$ points. It represents a purely-$J_2$ problem with no $J_1$ couplings. Ground states at this point are formed by the four $J_2$-sublattices ordering independently as ferromagnets. 

\begin{figure}
    \centering
    \includegraphics[width=0.8\linewidth]{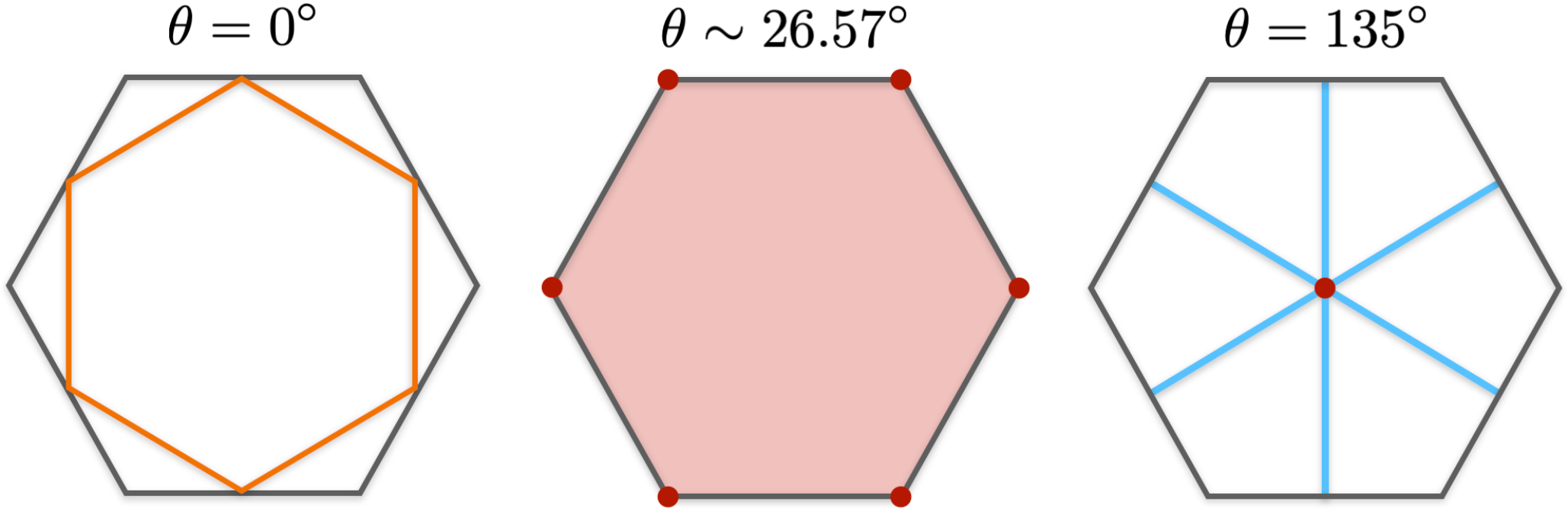}
    \caption{LT manifolds at three phase boundaries, shown in the projected Brillouin zone. At $\theta \sim 26.57^\circ$, we have a surface that covers the projected Brillouin zone. In addition, we have lines that are parallel to the stacking axis -- shown as dots. At $\theta=135^\circ$, we have lines within the projection plane as well as a line along the stacking axis (shown as a dot).  }
    \label{fig.boundaries}
\end{figure}

The LT minima at the other three boundaries, shown in Fig.~\ref{fig.boundaries}, all allow for extensive degeneracies. At $\theta = 135^\circ$, we have a line parallel to the stacking axis. This signifies that each layer may order independently as a ferromagnet. At $\theta \sim 26.57^\circ$, we have lines at the corners that are parallel to the stacking axis. These signify that each layer can develop 120$^\circ$ order independently. This boundary also hosts a surface of minima which contributes to degeneracy. Figures and details are provided in the supplementary material~\cite{supplementary}.

\begin{figure}
    \centering
    \includegraphics[width=0.5\linewidth]{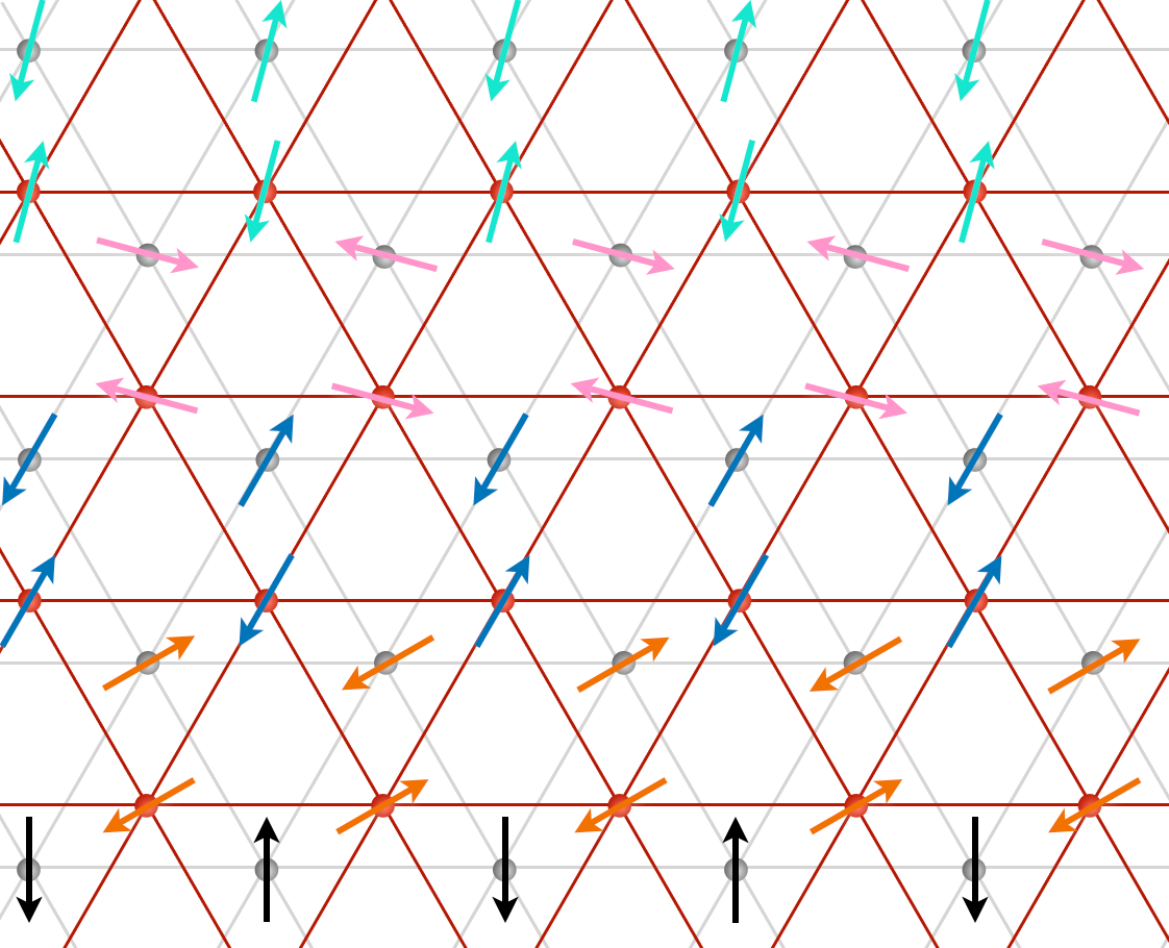}
    \caption{Extensive degeneracy in the purely-$J_1$ antiferromagnet at $\theta=0^\circ$. To construct a ground state, we start with a layer. Each line of this layer orders antiferromagnetically, with an independent N\'eel vector. On the next layer, we choose lines that run in the same direction and order them antiferromagnetically. N\'eel vectors are chosen such that each spin is anti-aligned with one spin of the previous layer. This construction can be extended to any number of layers with any stacking sequence. The degeneracy amounts to choosing an independent N\'eel vector for each line within any one layer. }
    \label{fig.theta0}
\end{figure}

The boundary at $\theta=0^\circ$ is particularly interesting with straight lines that do not pass through the $\Gamma$ point. On the fcc lattice, they encode the degeneracy discussed by Anderson in 1950: we may have independent N\'eel ordering on each plane of the cubic lattice \cite{Anderson1950}. The same extensive degeneracy exists on all close-packed lattices. As described in Fig.~\ref{fig.theta0}, we may independently choose N\'eel vectors on lines within a single plane. This fixes lines on successive layers. This construction gives rise to parallel surfaces (flat planes on fcc, corrugated sheets on all other structures) that are independently N\'eel-ordered.

\noindent{\color{blue} \it{Discussion}.}---Our central result is that an infinite family of magnets are equivalent at the level of the classical ground state. The arguments above are presented for magnets with Heisenberg couplings. However, the phase diagram of Fig.~\ref{fig.phasediagram} also holds for XY and Ising models. This follows from the fact that all four phases admit coplanar as well as collinear ground states -- see Figs.~\ref{fig.singleQ} and~\ref{fig.multiQ}. The underlying symmetry also holds in magnets that are not necessarily close-packed, but are composed of stacked triangular layers. Examples include delafossites\cite{Berthelot2011}, Na$_2$MnTeO$_6$\cite{Kurbakov2022,Cava2023} and rare-earth chalcogenides\cite{Xie2024} -- all known to have structural polytopes with stacking variations. While their structures are described as Barlow stackings, they carry differing in-plane and out-of-plane $J_1$ couplings. The LT matrix of Eq.~\ref{eq.Mmatrix} can be adapted to these materials by simply scaling the diagonal terms. The unitary transformation of Eq.~\ref{eq.unitary} continues to apply. This will result in a single phase diagram for all stacking polytopes. 

The gauge symmetry of Eq.~\ref{eq.unitary} can be broken by several features: couplings beyond $J_2$, spin wave fluctuations, directional Dzyaloshinskii-Moriya couplings, etc. Precise future studies can examine the degree of symmetry breaking in each close-packed structure. In the Supplementary Material\cite{supplementary}, we give an example of how spin-wave zero-point energy differs from one close-packing to another. An exciting future direction is to examine more complex frustrated magnets with stacked geometries. A prominent example is the diamond lattice $J_1-J_2$ model\cite{Bergman2007} which describes a `spiral liquid' phase in MnSc$_2$S$_4$. The diamond lattice has many stacking variants, which may host the same spiral surface if a gauge symmetry exists. However, quantum/thermal order-by-disorder may vary across polytopes.

\acknowledgments
We thank Jeffrey Rau and Connor Wilson for helpful discussions. SS thanks the Natural Sciences and Engineering Research Council of Canada (NSERC) for an Undergraduate Student Research Award. This work was supported by NSERC through Discovery Grant 2022-05240 (RG).

\bibliographystyle{apsrev4-2}
\bibliography{Refs}

@incollection{Cava2011,
  author    = {Cava, Robert J. and Holman, Katharine L. and McQueen, Tyrel and Welsh, Eric J. and West, D. Vincent and Williams, Anthony J.},
  title     = {The Geometries of Triangular Magnetic Lattices},
  booktitle = {Introduction to Frustrated Magnetism},
  editor    = {Lacroix, Claudine and Mendels, Philippe and Mila, Fr{\'e}d{\'e}ric},
  series    = {Springer Series in Solid-State Sciences},
  volume    = {164},
  publisher = {Springer Berlin Heidelberg},
  address   = {Berlin, Heidelberg},
  year      = {2011},
  chapter   = {6},
  pages     = {135--154},
  doi       = {10.1007/978-3-642-10589-0_6}
}

@article{Bergman2007,
	author = {Bergman, Doron and Alicea, Jason and Gull, Emanuel and Trebst, Simon and Balents, Leon},
	da = {2007/07/01},
	doi = {10.1038/nphys622},
	id = {Bergman2007},
	isbn = {1745-2481},
	journal = {Nature Physics},
	number = {7},
	pages = {487--491},
	title = {Order-by-disorder and spiral spin-liquid in frustrated diamond-lattice antiferromagnets},
	ty = {JOUR},
	url = {https://doi.org/10.1038/nphys622},
	volume = {3},
	year = {2007}}

@article{shull1951neutron,
  title={Neutron diffraction by paramagnetic and antiferromagnetic substances},
  author={Shull, C Gi and Strauser, WA and Wollan, EO},
  journal={Physical Review},
  volume={83},
  number={2},
  pages={333},
  year={1951},
  publisher={APS}
}

@article{Anderson1950,
  title = {Generalizations of the {W}eiss Molecular Field Theory of Antiferromagnetism},
  author = {Anderson, P. W.},
  journal = {Phys. Rev.},
  volume = {79},
  issue = {4},
  pages = {705--710},
  numpages = {0},
  year = {1950},
  month = {Aug},
  publisher = {American Physical Society},
  doi = {10.1103/PhysRev.79.705},
  url = {https://link.aps.org/doi/10.1103/PhysRev.79.705}
}

@article{luttinger1946theory,
  title={Theory of dipole interaction in crystals},
  author={Luttinger, JM and Tisza, L},
  journal={Physical Review},
  volume={70},
  number={11-12},
  pages={954},
  year={1946},
  publisher={APS}
}

@article{Luttinger1951,
  title = {A Note on the Ground State in Antiferromagnetics},
  author = {Luttinger, J. M.},
  journal = {Phys. Rev.},
  volume = {81},
  issue = {6},
  pages = {1015--1018},
  numpages = {0},
  year = {1951},
  month = {Mar},
  publisher = {American Physical Society},
  doi = {10.1103/PhysRev.81.1015},
  url = {https://link.aps.org/doi/10.1103/PhysRev.81.1015}
}

@article{Ziman1953,
doi = {10.1088/0370-1298/66/1/313},
url = {https://doi.org/10.1088/0370-1298/66/1/313},
year = {1953},
month = {jan},
publisher = {},
volume = {66},
number = {1},
pages = {89},
author = {J M Ziman},
title = {Antiferromagnetism by the Spin Wave Method {III}: Application to more Complex Systems},
journal = {Proceedings of the Physical Society. Section A}
}

@article{Lines1962,
    author = {Ter Haar, D. and Lines, M. E.},
    title = {A spin-wave theory of anisotropic antiferromagnetics},
    journal = {Philosophical Transactions of the Royal Society of London, Series A: Mathematical and Physical Sciences},
    volume = {255},
    number = {1049},
    pages = {1-30},
    year = {1962},
    month = {08},
    issn = {0080-4614},
    doi = {10.1098/rsta.1962.0008},
    url = {https://doi.org/10.1098/rsta.1962.0008}}

@article{Lines1964,
  title = {Green Functions in the Theory of Antiferromagnetism},
  author = {Lines, M. E.},
  journal = {Phys. Rev.},
  volume = {135},
  issue = {5A},
  pages = {A1336--A1346},
  numpages = {0},
  year = {1964},
  month = {Aug},
  publisher = {American Physical Society},
  doi = {10.1103/PhysRev.135.A1336},
  url = {https://link.aps.org/doi/10.1103/PhysRev.135.A1336}
}

@article{Oitmaa2023,
  title = {Ordered phases in the frustrated fcc lattice antiferromagnet},
  author = {Oitmaa, J.},
  journal = {Phys. Rev. B},
  volume = {108},
  issue = {1},
  pages = {014414},
  numpages = {5},
  year = {2023},
  month = {Jul},
  publisher = {American Physical Society},
  doi = {10.1103/PhysRevB.108.014414},
  url = {https://link.aps.org/doi/10.1103/PhysRevB.108.014414}
}

@article{Yamamoto1972,
author = {Yamamoto ,Yuji and Nagamiya ,Takeo},
title = {Spin Arrangements in Magnetic Compounds of the Rocksalt Crystal Structure},
journal = {Journal of the Physical Society of Japan},
volume = {32},
number = {5},
pages = {1248-1261},
year = {1972},
doi = {10.1143/JPSJ.32.1248},
URL = {     
        https://doi.org/10.1143/JPSJ.32.1248
}
}

@article{Oguchi1985,
author = {Oguchi ,Takehiko and Nishimori ,Hidetoshi and Taguchi ,Yoshihiro},
title = {The Spin Wave Theory in Antiferromagnetic {H}eisenberg Model on Face Centered Cubic Lattice},
journal = {Journal of the Physical Society of Japan},
volume = {54},
number = {12},
pages = {4494-4497},
year = {1985},
doi = {10.1143/JPSJ.54.4494},
URL = { 
        https://doi.org/10.1143/JPSJ.54.4494
}
}

@article{Henley1987,
    author = {Henley, Christopher L.},
    title = {Ordering by disorder: Ground‐state selection in fcc vector antiferromagnets},
    journal = {Journal of Applied Physics},
    volume = {61},
    number = {8},
    pages = {3962-3964},
    year = {1987},
    month = {04},
    issn = {0021-8979},
    doi = {10.1063/1.338570},
    url = {https://doi.org/10.1063/1.338570}
}

@article{Sun2018,
title = {The $J_1$-$J_2$ model on the face-centered-cubic lattices},
journal = {Journal of Magnetism and Magnetic Materials},
volume = {454},
pages = {176-184},
year = {2018},
issn = {0304-8853},
doi = {https://doi.org/10.1016/j.jmmm.2018.01.076},
url = {https://www.sciencedirect.com/science/article/pii/S030488531732824X},
author = {Ning-Ning Sun and Huai-Yu Wang}
}

@article{Zhitomirsky2020,
  title = {Quantum versus thermal fluctuations in the fcc antiferromagnet: Alternative routes to order by disorder},
  author = {Schick, R. and Ziman, T. and Zhitomirsky, M. E.},
  journal = {Phys. Rev. B},
  volume = {102},
  issue = {22},
  pages = {220405},
  numpages = {6},
  year = {2020},
  month = {Dec},
  publisher = {American Physical Society},
  doi = {10.1103/PhysRevB.102.220405},
  url = {https://link.aps.org/doi/10.1103/PhysRevB.102.220405}
}

@article{Balla2020,
  title = {Degenerate manifolds, helimagnets, and multi-{Q} chiral phases in the classical {H}eisenberg antiferromagnet on the face-centered-cubic lattice},
  author = {Balla, P\'eter and Iqbal, Yasir and Penc, Karlo},
  journal = {Phys. Rev. Res.},
  volume = {2},
  issue = {4},
  pages = {043278},
  numpages = {25},
  year = {2020},
  month = {Nov},
  publisher = {American Physical Society},
  doi = {10.1103/PhysRevResearch.2.043278},
  url = {https://link.aps.org/doi/10.1103/PhysRevResearch.2.043278}
}

@article{Zhitomirsky2022,
  title = {Ground-state selection by magnon interactions in a fcc antiferromagnet},
  author = {Schick, R. and G\"otze, O. and Ziman, T. and Zinke, R. and Richter, J. and Zhitomirsky, M. E.},
  journal = {Phys. Rev. B},
  volume = {106},
  issue = {9},
  pages = {094431},
  numpages = {10},
  year = {2022},
  month = {Sep},
  publisher = {American Physical Society},
  doi = {10.1103/PhysRevB.106.094431},
  url = {https://link.aps.org/doi/10.1103/PhysRevB.106.094431}
}

@article{Oja1993,
  title = {Long-range order produced by the interaction between spin waves in classical fcc {H}eisenberg models},
  author = {Heinil\"a, Marko T. and Oja, Aarne S.},
  journal = {Phys. Rev. B},
  volume = {48},
  issue = {22},
  pages = {16514--16523},
  numpages = {0},
  year = {1993},
  month = {Dec},
  publisher = {American Physical Society},
  doi = {10.1103/PhysRevB.48.16514},
  url = {https://link.aps.org/doi/10.1103/PhysRevB.48.16514}
}

@article{Diep1992,
  title = {First-order transition in the hexagonal-close-packed lattice with vector spins},
  author = {Diep, H. T.},
  journal = {Phys. Rev. B},
  volume = {45},
  issue = {6},
  pages = {2863--2867},
  numpages = {0},
  year = {1992},
  month = {Feb},
  publisher = {American Physical Society},
  doi = {10.1103/PhysRevB.45.2863},
  url = {https://link.aps.org/doi/10.1103/PhysRevB.45.2863}
}

@article{Diep2012,
  title = {Hexagonal-close-packed lattice: Ground state and phase transition},
  author = {Hoang, Danh-Tai and Diep, H. T.},
  journal = {Phys. Rev. E},
  volume = {85},
  issue = {4},
  pages = {041107},
  numpages = {6},
  year = {2012},
  month = {Apr},
  publisher = {American Physical Society},
  doi = {10.1103/PhysRevE.85.041107},
  url = {https://link.aps.org/doi/10.1103/PhysRevE.85.041107}
}

@article{Niggemann2020,
doi = {10.1088/1361-648X/ab4480},
url = {https://doi.org/10.1088/1361-648X/ab4480},
year = {2019},
month = {oct},
publisher = {IOP Publishing},
volume = {32},
number = {2},
pages = {024001},
author = {Niggemann, Nils and Hering, Max and Reuther, Johannes},
title = {Classical spiral spin liquids as a possible route to quantum spin liquids},
journal = {Journal of Physics: Condensed Matter}
}

@article{Aplesnin1989,
  author  = {Aplesnin, S. S. and Gekht, R. S.},
  title   = {Incommensurate and aperiodic structures in frustrated antiferromagnets with hcp
lattices},
  journal = {Soviet Physics JETP},
  volume  = {69},
  pages   = {1224--1230},
  year    = {1989},
  note    = {Translated from Zh. Eksp. Teor. Fiz. 96, 2163 (1989)}
}

@article{Hagg1943,
	author = {H\"agg, G.},
	journal = {Ark. Kem. Mineral. Geol.},
	pages = {1},
	ty = {JOUR},
	volume = {163},
	year = {1943}}

@article{supplementary,
journal={Supplemental~material}
}

@article{He2024,
  title = {Metallic Bonding in Close-Packed Structures: Structural Frustration from a Hidden Gauge Symmetry},
  author = {He, Eric and Wilson, C. M. and Ganesh, R.},
  journal = {Phys. Rev. Lett.},
  volume = {133},
  issue = {25},
  pages = {256401},
  numpages = {6},
  year = {2024},
  month = {Dec},
  publisher = {American Physical Society},
  doi = {10.1103/PhysRevLett.133.256401},
  url = {https://link.aps.org/doi/10.1103/PhysRevLett.133.256401}
}

@misc{KrishnaPandey1981,
	author = {{Krishna, P.} and {Pandey, D.}},
	note = {{I}nternational {U}nion of {C}rystallography},
	title = {Close-packed structures},
	url = {https://www.iucr.org/__data/assets/pdf_file/0015/13254/5.pdf},
year=1981
}

@article{Moessner_Chalker1998,
  title = {Low-temperature properties of classical geometrically frustrated antiferromagnets},
  author = {Moessner, R. and Chalker, J. T.},
  journal = {Phys. Rev. B},
  volume = {58},
  issue = {18},
  pages = {12049--12062},
  numpages = {0},
  year = {1998},
  month = {Nov},
  publisher = {American Physical Society},
  doi = {10.1103/PhysRevB.58.12049},
  url = {https://link.aps.org/doi/10.1103/PhysRevB.58.12049}
}

@article{Henley2005,
  title = {Power-law spin correlations in pyrochlore antiferromagnets},
  author = {Henley, C. L.},
  journal = {Phys. Rev. B},
  volume = {71},
  issue = {1},
  pages = {014424},
  numpages = {11},
  year = {2005},
  month = {Jan},
  publisher = {American Physical Society},
  doi = {10.1103/PhysRevB.71.014424},
  url = {https://link.aps.org/doi/10.1103/PhysRevB.71.014424}
}

@article{Henley2010,
   author = "Henley, Christopher L.",
   title = "The “Coulomb Phase” in Frustrated Systems", 
   journal= "Annual Review of Condensed Matter Physics",
   year = "2010",
   volume = "1",
   number = "Volume 1, 2010",
   pages = "179-210",
   doi = "https://doi.org/10.1146/annurev-conmatphys-070909-104138",
   url = "https://www.annualreviews.org/content/journals/10.1146/annurev-conmatphys-070909-104138",
   publisher = "Annual Reviews",
   issn = "1947-5462",
   type = "Journal Article",
  }

@article{Litvin1974,
title = {The Luttinger-Tisza method},
journal = {Physica},
volume = {77},
number = {2},
pages = {205-219},
year = {1974},
issn = {0031-8914},
doi = {https://doi.org/10.1016/0031-8914(74)90257-2},
url = {https://www.sciencedirect.com/science/article/pii/0031891474902572},
author = {D.B. Litvin}
}

@article{Pandey2020,
  title = {Analytical theory of pyrochlore cooperative paramagnets},
  author = {Pandey, Akshat and Moessner, Roderich and Castelnovo, Claudio},
  journal = {Phys. Rev. B},
  volume = {101},
  issue = {11},
  pages = {115107},
  numpages = {7},
  year = {2020},
  month = {Mar},
  publisher = {American Physical Society},
  doi = {10.1103/PhysRevB.101.115107},
  url = {https://link.aps.org/doi/10.1103/PhysRevB.101.115107}
}

@article{Canals1998,
  title = {Pyrochlore Antiferromagnet: A Three-Dimensional Quantum Spin Liquid},
  author = {Canals, B. and Lacroix, C.},
  journal = {Phys. Rev. Lett.},
  volume = {80},
  issue = {13},
  pages = {2933--2936},
  numpages = {0},
  year = {1998},
  month = {Mar},
  publisher = {American Physical Society},
  doi = {10.1103/PhysRevLett.80.2933},
  url = {https://link.aps.org/doi/10.1103/PhysRevLett.80.2933}
}

@article{Reimers1992,
  title = {Absence of long-range order in a three-dimensional geometrically frustrated antiferromagnet},
  author = {Reimers, J. N.},
  journal = {Phys. Rev. B},
  volume = {45},
  issue = {13},
  pages = {7287--7294},
  numpages = {0},
  year = {1992},
  month = {Apr},
  publisher = {American Physical Society},
  doi = {10.1103/PhysRevB.45.7287},
  url = {https://link.aps.org/doi/10.1103/PhysRevB.45.7287}
}

@article{Khatua_quadrumer,
  title = {Quantum spin quadrumer},
  author = {Khatua, Subhankar and Shankar, R. and Ganesh, R.},
  journal = {Phys. Rev. B},
  volume = {97},
  issue = {5},
  pages = {054403},
  numpages = {13},
  year = {2018},
  month = {Feb},
  publisher = {American Physical Society},
  doi = {10.1103/PhysRevB.97.054403},
  url = {https://link.aps.org/doi/10.1103/PhysRevB.97.054403}
}

@book{Szpiro2003kepler,
  title={Kepler's Conjecture},
  author={Szpiro, G.G.},
  year={2003},
  publisher={John Wiley}
}

@article{Barrett1955,
author = {Barrett, C. S.},
title = {??},
journal = {J. Inst. Metals},
volume = {74},
pages = {43},
year={1955}
}

@article{barrett_PR_1947,
  title = {A Low Temperature Transformation in Lithium},
  author = {Barrett, Charles S.},
  journal = {Phys. Rev.},
  volume = {72},
  issue = {3},
  pages = {245--245},
  numpages = {0},
  year = {1947},
  month = {Aug},
  publisher = {American Physical Society},
  doi = {10.1103/PhysRev.72.245},
  url = {https://link.aps.org/doi/10.1103/PhysRev.72.245}
}

@article{ackland_Science_2017,
  author = {Graeme J. Ackland  and Mihindra Dunuwille  and Miguel Martinez-Canales  and Ingo Loa  and Rong Zhang  and Stanislav Sinogeikin  and Weizhao Cai  and Shanti Deemyad },
  title = {Quantum and isotope effects in lithium metal},
  journal = {Science},
  volume = {356},
  number = {6344},
  pages = {1254-1259},
  year = {2017},
  doi = {10.1126/science.aal4886},
  URL = {https://www.science.org/doi/abs/10.1126/science.aal4886}
}

@article{Berthelot2011,
    author = {Berthelot, R. and Pollet, M. and Doumerc, J.-P. and Delmas, C.},
    title = {First Experimental Evidence of a New D4-AgCoO2 Delafossite Stacking},
    journal = {Inorganic Chemistry},
    volume = {50},
    number = {10},
    pages = {4529-4536},
    year = {2011},
    month = {04},
    issn = {0020-1669},
    doi = {10.1021/ic2002354},
    url = {https://doi.org/10.1021/ic2002354}
}

@article{Cava2023,
  title = {Spin disorder in a stacking polytype of a layered magnet},
  author = {Xu, Xianghan and Cheng, Guangming and Ni, Danrui and Gui, Xin and Xie, Weiwei and Yao, Nan and Cava, R. J.},
  journal = {Phys. Rev. Mater.},
  volume = {7},
  issue = {2},
  pages = {024407},
  numpages = {8},
  year = {2023},
  month = {Feb},
  publisher = {American Physical Society},
  doi = {10.1103/PhysRevMaterials.7.024407},
  url = {https://link.aps.org/doi/10.1103/PhysRevMaterials.7.024407}
}

@article{Kurbakov2022,
  title = {Commensurate helicoidal order in the triangular layered magnet ${\mathrm{Na}}_{2}\mathrm{MnTe}{\mathrm{O}}_{6}$},
  author = {Kurbakov, A. I. and Susloparova, A. E. and Pomjakushin, V. Y. and Skourski, Y. and Vavilova, E. L. and Vasilchikova, T. M. and Raganyan, G. V. and Vasiliev, A. N.},
  journal = {Phys. Rev. B},
  volume = {105},
  issue = {6},
  pages = {064416},
  numpages = {9},
  year = {2022},
  month = {Feb},
  publisher = {American Physical Society},
  doi = {10.1103/PhysRevB.105.064416},
  url = {https://link.aps.org/doi/10.1103/PhysRevB.105.064416}
}

@article{Xie2024,
doi = {10.1088/0256-307X/41/11/117505},
url = {https://doi.org/10.1088/0256-307X/41/11/117505},
year = {2024},
month = {dec},
publisher = {Chinese Physical Society and IOP Publishing Ltd},
volume = {41},
number = {11},
pages = {117505},
author = {Xie, Mingtai and Zhuo, Weizhen and Cai, Yanzhen and Zhang, Zheng and Zhang, Qingming},
title = {Rare-Earth Chalcogenides: An Inspiring Playground for Exploring Frustrated Magnetism},
journal = {Chinese Physics Letters}
}

@book{auerbach2012interacting,
  title={Interacting electrons and quantum magnetism},
  author={Auerbach, Assa},
  year={2012},
  publisher={Springer Science \& Business Media}
}

\newpage

\newpage

\onecolumngrid 
\setcounter{secnumdepth}{3}
\renewcommand{\theequation}{S\arabic{equation}}
\renewcommand{\thefigure}{S\arabic{figure}}
\begin{center}
 \textbf{\large Supplemental Material for ``Local-moment magnets on all close-packed lattices are equivalent''}\\[.5cm]
Sam Studdy$^1$, Alwyn Jose Raja$^2$ and R. Ganesh$^1$\\[.4cm]
{\itshape ${}^2$Department of Physics, Brock University, St. Catharines, Ontario L2S 3A1, Canada\\
\itshape ${}^1$Department of Physics, Indian Institute of Technology Madras, Chennai 600036, India\\
}

(Dated: \today)\\[2cm]
\end{center}

\section{The Luttinger-Tisza matrix in a close-packed magnet}
In classical Heisenberg magnets with lattice translational symmetry, the LT approach first converts the Hamiltonian to momentum space, 
\begin{eqnarray}
    H = \sum_{ij} J_{ij} \mathbf{S}_i \cdot \mathbf{S}_j \rightarrow \sum_{\mathbf{q}} J_{\alpha\beta} (\mathbf{q}) \mathbf{S}_{-\mathbf{q},\alpha} \cdot \mathbf{S}_{\mathbf{q},\beta}.
\end{eqnarray}
Here, $\mathbf{q}$ represents a point within the Brillouin zone. The indices $\alpha$ and $\beta$ represent sublattices. In the unit cell described in the main text and Fig.~\ref{fig.unitcell}, they run from 1 to $M$. The $()\cdot()$ operation denotes a dot product between two Heisenberg spins. In XY or Ising models, this operation is to be modified suitably. Lattice translational symmetry ensures that the Hamiltonian conserves momentum. To have a legitimate spin configuration, we must have a fixed spin length $S$ at every site. In the LT approach, the spin-length constraint is replaced by a weaker constraint that fixes the `average' spin length. This is implemented by a Lagrange multiplier,
\begin{eqnarray}
H \rightarrow H - \lambda \sum_{\mathbf{q},\alpha} \mathbf{S}_{-\mathbf{q},\alpha}\cdot \mathbf{S}_{\mathbf{q},\alpha}.    
\end{eqnarray}
The Hamiltonian, with the Lagrange multiplier, can then be written as a quadratic form,
\begin{eqnarray}
    H = \sum_{\mathbf{q}} \sum_u \left(
    \begin{array}{cccc} 
    S_{-\mathbf{q},1}^u &     S_{-\mathbf{q},2}^u & \ldots &     S_{-\mathbf{q},M}^u \end{array}\right)
    M(\mathbf{q})\left(\begin{array}{c}
    S_{\mathbf{q},1}^u \\     S_{\mathbf{q},2}^u \\ \vdots \\     S_{\mathbf{q},M}^u \end{array}\right),
\end{eqnarray}
where the index $u$ runs over all components of the spin vector. For Heisenberg spins, $u$ runs over $x,y,z$. As the spin components at any given site are real-valued, we have $S_{-\mathbf{q},\alpha}^u = \left(S_{\mathbf{q},\alpha}^u\right)^*$. We next minimize the energy, treating ${S}_{\mathbf{q},\alpha}^u$ as independent variables. To solve this problem, we may go to the eigenbasis of $M(\mathbf{q})$. We identify the lowest eigenvalue, over all $\mathbf{q}$, as a lower bound for the energy of the system.

This lower bound can be saturated if we are able to construct a legitimate spin state using only the eigenvector corresponding to the minimum eigenvalue. We assert that this can be achieved if the eigenvector consists of pure phases. We can then assign spin orientations within a plane, say the XY plane in spin space. This results in a legitimate coplanar spin arrangement that will saturate the LT bound. Expressing each element of the LT eigenvector as $e^{i\varphi_\alpha} = \cos \varphi_\alpha + i \sin\varphi_\alpha$, we may assign an orientation to spin-$\alpha$ of the unit cell as follows. We identify the cosine as the X component of spin-$\alpha$ and the sine as its Y component (assuming spin length is set to unity). The change in orientation from one unit cell to the next is captured by the phase shift $\mathbf{q}_{min}\cdot \mathbf{r}$, where $\mathbf{q}_{min}$ is the momentum that carries the minimum eigenvalue and $\mathbf{r}$ is the displacement between unit cells. When Fourier transformed, this state will produce non-zero Fourier components exclusively at $\mathbf{q}_{min}$, producing the same ground state energy as the LT lower bound. 
We describe this construction with an explicit example below. 

\subsection{The LT matrix for the dhcp lattice}
As an example, we consider the dhcp (double hexagonal close-packed) structure shown in Fig.~\ref{fig.unitcell}(b). The unit cell contains four sites, one from each layer of the ABAC arrangement. The LT matrix is naturally $4 \times 4$. The diagonal entries encode couplings from each sublattice to itself, i.e., couplings that are purely intra-layer. Each site has six $J_1$ neighbours in the same plane, located at $\pm \hat{a},\pm\hat{b}$ and $\pm(\hat{a}+\hat{b})$, where $\hat{a}$ and $\hat{b}$ are the primitive lattice vectors shown in Figs.~\ref{fig.unitcell}(a,b). With these six bonds, each diagonal entry takes the form given by Eq.~\ref{eq.Vdiagonal}. 

We next consider the $(1,2)$ element, representing the coupling from the base layer, A, to the B layer that is directly above. Each site in the A layer has six neighbours in the adjacent B layer -- three $J_1$ neighbours and three $J_2$ neighbours. The $J_1$ neighbours are located at displacements of $\hat{c}/4 + \mathbf{u}$, where $\mathbf{u}$ are the three in-plane vectors defined below Eq.~\ref{eq.phidef} of the main text. The $J_2$ neighbours are located at $\hat{c}/4 + \mathbf{v}$, where $\mathbf{v}$ are also in-plane vectors defined below Eq.~\ref{eq.phidef}. With these couplings, the $(1,2)$ element of the LT matrix takes the form,
\begin{equation}
    M_{1,2}(\mathbf{q}) = \frac{e^{i q_c/4}}{2} \Big[
     J_1 \sum_{\mathbf{u}} e^{i\mathbf{q}\cdot\mathbf{u}} + J_2 \sum_{\mathbf{v}} e^{i\mathbf{q}\cdot\mathbf{v}}
    \Big].
    \label{eq.M12}
\end{equation}
This is precisely the expression given in Eq.~\ref{eq.Vj} with $M=4$ and $\sigma_j = 1$. The $(2,3)$ element can be found by similar arguments. The only difference is the $\mathbf{u}$ and $\mathbf{v}$ vectors must be inverted as we move from layer B to the A layer above, corresponding to the H\"agg code entry of $-1$,
\begin{equation}
    M_{2,3}(\mathbf{q}) = \frac{e^{i q_c/4}}{2} \Big[
     J_1 \sum_{\mathbf{u}} e^{-i\mathbf{q}\cdot\mathbf{u}} + J_2 \sum_{\mathbf{v}} e^{-i\mathbf{q}\cdot\mathbf{v}}
    \Big].
    \label{eq.M23}
\end{equation}
Entries such as $(1,3)$ vanish as there are no couplings from the first to the third layer.  
Proceeding in this manner, we obtain
\begin{equation}    M(\vec{k})=\begin{pmatrix}
        V_D&V_+&0&V_+^*\\
        V_+^*&V_D&V_-&0\\
        0&V_-^*&V_D&V_-\\
        V_+&0&V_-^*&V_D
    \end{pmatrix},
\end{equation}
where $V_+$, $V_-$ and $V_D$ are all functions of $\mathbf{q}$. Here, $V_+$ and $V_-$ are the same as $M_{1,2}(\mathbf{q})$ and $M_{2,3}(\mathbf{q})$ above. 
The arrangement of $V_+$'s and $V_-$'s in the right-of-diagonal reflects the dhcp structure as expressed in H\"agg code notation, $(+--+)$. 

To demonstrate the pure-phases condition, we consider the dhcp lattice at a particular point on the phase diagram, say in the $W$ phase, for example $J_1=1$, $J_2=\frac{1}{10}$. In this region, the LT minima occur at the $W$ points in the projected Brillouin zone. The minimum eigenvalue is $E_{min} = -2J_1 + J_2$, which we identify as the LT lower bound for the classical ground state energy. For concreteness, consider the $W$ point with coordinates $(q_x=\pi,q_y=0)$ in the projected hexagonal Brillouin zone (in units where the lattice constant $\vert \hat{a} \vert = \vert \hat{b} \vert = 1)$. At this point, the eigenvector corresponding to the minimum eigenvalue is given by $\frac{1}{2}(-1,1,-1,1) = \frac{1}{2}(e^{i\pi},e^{i\cdot0},e^{i\pi},e^{i\cdot0})$. As all entries have the same magnitude, this eigenvector satisfies the pure-phase condition. To construct a coplanar ground state with spins in the XY plane, we use the four phases in the eigenvector $\phi_j$, $j=1,\ldots,4$. The spin orientation at sublattice $j$ is taken to be $\phi_j+\mathbf{W}\cdot\mathbf{r}_j$, where $\mathbf{r}_j$ is the position of the $j^\mathrm{th}$ site within the unit cell. In the reference unit cell, we obtain the A, B, A and C spins (one in each layer) to be $(-\hat{x},\hat{y},-\hat{x},\hat{x})$. In other unit cells, spin orientations acquire an additional phase of $\mathbf{W}\cdot \mathbf{r} = \pi r_x$, where $\mathbf{r}$ is the separation from the reference unit cell. This produces a minimum-energy spin configuration, with the first two layers as shown in Fig.~\ref{fig.singleQ}(d).

\section{An explicit example of the unitary transformation }
In the main text, Eq.~\ref{eq.unitary} presents a unitary transformation that connects the LT matrices of two distinct close-packed structures. We present an explicit example, connecting (ABAB) and (ABAC) structures. The LT matrices for these structures are given by
\begin{equation*}
    M_{ABAC}(\vec{k})=\begin{pmatrix}
        V_D&V_+&0&V_+^*\\
        V_+^*&V_D&V_-&0\\
        0&V_-^*&V_D&V_-\\
        V_+&0&V_-^*&V_D
    \end{pmatrix},~~~
    M_{ABAB}(\vec{k})=\begin{pmatrix}
        V_D&V_+&0&V_-^*\\
        V_+^*&V_D&V_-&0\\
        0&V_-^*&V_D&V_+\\
        V_-&0&V_+^*&V_D
    \end{pmatrix},
\end{equation*}
where the entries are functions of momentum $\mathbf{q}$. We have $    V_\pm=e^{i\frac{q_c}{4}}f(\pm\mathbf{q}_\parallel)$, where $f(\mathbf{q}_\parallel)$ is defined in Eq.~\ref{eq.phidef} of the main text. 

We consider a transformation matrix $\mathcal{U}=\mathrm{Diag}(e^{i\theta_1},e^{i\theta_2},e^{i\theta_3},e^{i\theta_4})$. We perform a similarity transform on $M_{\text{ABAC}}$ using this matrix and demand that the result match $M_{\text{ABAB}}$,
\begin{eqnarray}
\begin{pmatrix}
        V_D&V_+e^{i(\theta_2-\theta_1)}&0&V_+^*e^{i(\theta_4-\theta_1)}\\
        V_+^*e^{-i(\theta_2-\theta_1)}&V_D&V_-e^{i(\theta_3-\theta_2)}&0\\
        0&V_-^*e^{i(\theta_2-\theta_3)}&V_D&V_-e^{i(\theta_4-\theta_3)}\\
        V_+e^{i(\theta_1-\theta_4)}&0&V_-^*e^{i(\theta_3-\theta_4)}&V_D\\
    \end{pmatrix}
    =\label{Unitary transformation condition}\begin{pmatrix}
        V_D&V_+&0&V_-^*\\
        V_+^*&V_D&V_-&0\\
        0&V_-^*&V_D&V_+\\
        V_-&0&V_+^*&V_D
    \end{pmatrix}.
    \label{eq.matrixequality}
\end{eqnarray}
By inspection, we have $\theta_1=\theta_2=\theta_3$. From the $(3,4)$ element, we have $V_- e^{i(\theta_4-\theta_3)}=V_+$. To fix $\theta$, we may use Eq.~\ref{eq.phidef} of the main text which yields $V_-=V_+e^{-2i\phi(\mathbf{q}_\parallel)}$. We may satisfy Eq.~\ref{eq.matrixequality} by setting $\theta_1 = \theta_2 = \theta_3 = 0, \theta_4 = 2\phi(\mathbf{q}_\parallel)$. 

\section{Gauge symmetry as the Aharanov-Bohm effect in layer space}
The LT matrix in Eq.~\ref{eq.Mmatrix} of the main text can be interpreted as follows. For a given structure, at each $\mathbf{q}_\parallel$ and $q_c$, we have an $M\times M$ Hermitian matrix. This can be viewed as an effective tight binding model describing a particle that moves on a closed ring of $M$ sites. With the tridiagonal nature of the matrix, the particle can only hop between nearest neighbours. When hopping from site $j$ to site $j+1$, the particle picks up a phase given 
\begin{eqnarray}
    \Phi_{j,j+1} = \frac{q_c}{M} + \phi(\mathbf{q}_\parallel) ~\sigma_j.
\end{eqnarray}
Here, $\phi(\mathbf{q}_\parallel)$ is a function of $ \mathbf{q}_\parallel $ as well as the coupling strengths $J_1$ and $J_2$. The phase picked up by the particle is reminiscent of the Aharonov-Bohm effect where the wavefunction of a charged particle picks up a phase in the presence of a magnetic field. In this picture, the phase associated with each hopping process is given by the Peierls substitution formula. It is the line integral of the magnetic vector potential, $\Phi_{j,j+1} = \int_j^{j+1} \mathbf{A} \cdot \mathbf{dl}$. This phase is clearly gauge-dependent as the vector potential is not uniquely defined.

The eigenvalues of the matrix are stationary-state energies of the effective particle. As energies are observable quantities, they cannot be gauge-dependent. It follows that the eigenvalues of the matrix cannot depend on each hopping-phase. Rather, they can only depend on the net phase obtained upon going around the ring, $\Phi_{net} = \oint \mathbf{A}\cdot \mathbf{dl} =  \sum_{j=1}^M \Phi_{j,j+1}$. This net flux corresponds to the line integral of the vector potential around the ring, which in turn, corresponds to the net magnetic flux that pierces the ring. With the geometry of a single closed ring, this is the only gauge-invariant quantity that can be constructed from the phases. It follows that the eigenvalues of the matrix are set by
$\sum_{j=1}^M \Phi_{j,j+1} = q_c + \phi(q_\parallel) \sum_{j=1}^M \sigma_j$. 

We may now compare the LT matrices of two different structures: $\{ \sigma\}$ and $\{\chi\}$, at the same $\mathbf{q}_\parallel$ but at two different $q_c$ values, $q_{c,\sigma}$ and $q_{c,\chi}$. The eigenvalues of the two matrices are identical if they correspond to the same net flux in the effective particle-on-a-ring picture, i.e., if
\begin{eqnarray}
q_{c,\sigma} + \phi(\mathbf{q}_\parallel) \sum_{j=1}^M \sigma_j = q_{c,\chi} + \phi(\mathbf{q}_\parallel) \sum_{j=1}^M \chi_j.  
\label{eq.equalflux}
\end{eqnarray}
We define $\mathbf{Q}$ as the shift along $\hat{c}$ that is required to have the eigenvalues matched. Rearranging Eq.~\ref{eq.equalflux}, we obtain
\begin{eqnarray}
    Q = q_{c,\chi} - q_{c,\sigma} = \phi(\mathbf{q}_\parallel) \sum_{j=1}^M (\sigma_j - \chi_j).
 \end{eqnarray}
This is precisely the condition for unitary equivalence, matching the expression for $Q$ in Eq.~\ref{eq.Qdef} of the main text.

\begin{figure}
    \centering
    \includegraphics[width=0.3\linewidth]{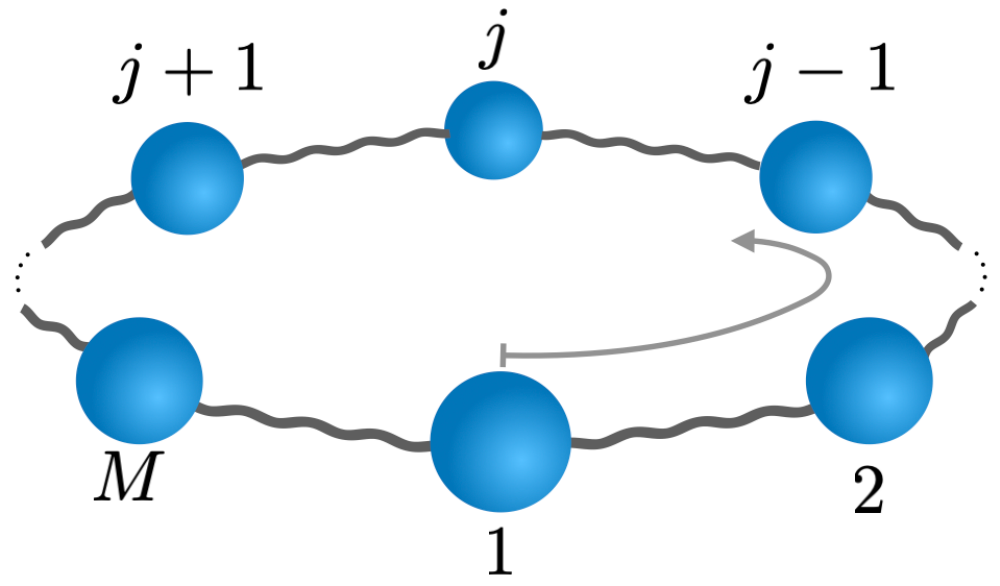}
    \caption{The effective hopping picture for the LT matrix of an $M$-layer close-packed magnet.}
    \label{fig.ring}
\end{figure}

\section{Multi-$Q$ states}

In Fig.~\ref{fig.phasediagram} of the main text, we describe the locations of the LT minima in each of the four phases. We now demonstrate how multiple minima can be combined into legitimate multi-$Q$ ground states.

\subsection{$(\Gamma,M)$ Phase}
The $(\Gamma,M)$ phase encompasses the $\theta=\pi/2$ point, where $J_1 =0,~J_2 > 0$. At this point, the four $J_2$ sublattices decouple from one another. With antiferromagnetic $J_2$ couplings, each sublattice develops N\'eel order independently. Any such configuration can be expressed as a superposition of four spirals: one at the $\Gamma$ and one at each of the three $M$ points. To see this, we consider spins on a single triangular layer. We fix the spin at position $\mathbf{r}_{m,n} = m\hat{a}+n\hat{b}$ as 
\begin{eqnarray}
    \mathbf{S}_{m,n} = \mathbf{u} + \sum_{j=1,2,3}\mathbf{v}_j e^{i\mathbf{M}_j \cdot \mathbf{r}_{m,n}}. 
\end{eqnarray}
Here, the vector $\mathbf{u}$ represents a spiral at the $\Gamma$ point. The three $\mathbf{v}_j$'s are components at each of the $M$ points. Each of these four vectors has three real-valued components. The real-valued character comes from the fact that the four momenta satisfy $\mathbf{Q}\equiv -\mathbf{Q}$ (up to a reciprocal lattice vector). As $\mathbf{S}_{-\mathbf{Q}} = \mathbf{S}_{\mathbf{Q}}^*$, the vectors must be real-valued. This ansatz naturally leads to a four-site unit cell within the layer, consistent with Fig.~\ref{fig.singleQ}(b). To ensure a legitimate spin configuration, we require the spin length to be unity at every site. This reduces to 
\begin{eqnarray}
    1 = \vert \mathbf{u} \vert^2 + \sum_{j=1,2,3} \vert \mathbf{v}_j \vert^2;~~~~\mathbf{u}\cdot \mathbf{v}_i + \epsilon_{ijk}\mathbf{v}_j\cdot\mathbf{v}_k = 0,~i=1,2,3.
    \label{eq.GammaMconstraints}
\end{eqnarray}
We have used $e^{i\vec{M}_j \cdot \mathbf{r}_{m,n}} = \pm1$ as $(m,n)$ are varied, as well as $e^{i\vec{M}_i \cdot \mathbf{r}_{m,n} }e^{i\vec{M}_j \cdot \mathbf{r}_{m,n}}=\delta_{i,j} + \epsilon_{i,j,k}e^{i\vec{M}_k \cdot \mathbf{r}_{m,n}}$. With four vectors, $\mathbf{u}, \mathbf{v}_{1-3}$, each with three components, we have twelve real-valued parameters. With the four constraints in Eq.~\ref{eq.GammaMconstraints}, we arrive at eight degrees of freedom. In the real-space picture, we have four independent N\'eel vectors. With two degrees of freedom to determine each N\'eel vector, we once again arrive at eight degrees of freedom.

To determine spin orientations in other layers, it suffices to ensure that the four $J_2$ sublattices are antiferromagnetically ordered. It is easy to see that, regardless of stacking sequence, this state saturates the LT bound for the classical ground state energy. The four-site motif in one layer repeats in every layer of the close-packed structure. 

\subsection{$M$ Phase}

We have three LT minima, one at each $M$ point of the projected Brillouin zone. We focus on a single layer and write a generic multi-$Q$ state as
\begin{eqnarray}
     \mathbf{S}_{m,n} = \sum_{j=1,2,3}\mathbf{v}_j e^{i\mathbf{M}_j \cdot \mathbf{r}_{m,n}}, 
\end{eqnarray}
where $\mathbf{r}_{m,n} = m\hat{a} + n\hat{b}$ is a generic site on the layer and $\mathbf{v}$'s are vectors with real-valued components. The real-valuedness arises from the fact that the $\mathbf{M}$ vectors satisfy $\mathbf{Q} \equiv -\mathbf{Q}$ (up to a reciprocal lattice vector). It is easy to see that this ansatz produces a repeating four-site motif within the layer. It is also easy to check that the four spins within the motif add to zero, as described in the main text. 

Demanding that the spin length at each site is unity, we arrive at three constraints,
\begin{eqnarray}
    1 = \sum_{j=1,2,3} \vert\mathbf{v}_j \vert^2;~~~ \mathbf{v}_1 \cdot \mathbf{v}_2 = \mathbf{v}_2 \cdot \mathbf{v}_3 = \mathbf{v}_3 \cdot \mathbf{v}_1 = 0.
    \label{eq.Mconstraints}
\end{eqnarray}
With three vectors, each with three components, we have nine degrees of freedom. With the four constraints in Eq.~\ref{eq.Mconstraints}, we arrive at five degrees of freedom. To determine spin orientations on other layers, we assert that each $J_2$-sublattice must be ferromagnetically ordered. It can be easily checked that this produces the same classical ground state energy as the LT minimum, regardless of stacking sequence. 

\subsection{$W$ Phase}

We have minima at the six $W$ points. Unlike the $\Gamma$ point and $M$ points, the $W$'s do not satisfy $\mathbf{Q}\equiv -\mathbf{Q}$. Rather, we have $\mathbf{W}_4 = -\mathbf{W}_1$, $\mathbf{W}_5 = -\mathbf{W}_2$ and $\mathbf{W}_6= -\mathbf{W}_3$, allowing us to group the $\mathbf{W}$'s into three pairs of opposite momenta. We write a generic multi-$Q$ state on a given layer as
\begin{eqnarray}
        \mathbf{S}_{m,n} = \mathbf{u} e^{i\mathbf{W}_1 \cdot \mathbf{r}_{m,n}} + \mathbf{u}^* e^{-i\mathbf{W}_1 \cdot \mathbf{r}_{m,n}} + 
        \mathbf{v} e^{i\mathbf{W}_2 \cdot \mathbf{r}_{m,n}} + \mathbf{v}^* e^{-i\mathbf{W}_2 \cdot \mathbf{r}_{m,n}} + 
        \mathbf{w} e^{i\mathbf{W}_3 \cdot \mathbf{r}_{m,n}} + \mathbf{w}^* e^{-i\mathbf{W}_3 \cdot \mathbf{r}_{m,n}}. 
\end{eqnarray}
We have used the requirement that $\mathbf{S}_{-\mathbf{Q}} = \mathbf{S}_{\mathbf{Q}}^*$ for spin components at all sites to be real valued. Here, $\mathbf{u}$, $\mathbf{v}$ and $\mathbf{w}$ are three-component complex-valued vectors. We separate them into real and imaginary parts, as $\mathbf{u}=\mathbf{u}_R + i \mathbf{u}_I, \mathbf{v}=\mathbf{v}_R + i \mathbf{v}_I, \mathbf{w}=\mathbf{w}_R + i \mathbf{w}_I$.

Upon demanding that the spin length at each site must be unity, we arrive at the following constraints:
\begin{enumerate}
    \item $\mathbf{u}\cdot\mathbf{u}^* + \mathbf{v}\cdot\mathbf{v}^* + \mathbf{w}\cdot\mathbf{w}^*=1/2$.
    \item $\mathbf{u}^2 + \left(\mathbf{u}^*\right)^2=\mathbf{v}^2+ \left(\mathbf{v}^*\right)^2=\mathbf{w}^2+ \left(\mathbf{w}^*\right)^2=0$. \item $\mathbf{u}\cdot\mathbf{v}^*=\mathbf{w}\cdot\mathbf{v}^*=\mathbf{u}\cdot\mathbf{w}^*=0$, with their conjugates also vanishing. 
\end{enumerate}
With some further simplifications using the fact that each $W$ momentum is one fourth of a reciprocal lattice vector, these constraints can be expressed as 
\begin{enumerate}
    \item $\mathbf{u}_R^2+\mathbf{u}_I^2+\mathbf{v}_R^2+\mathbf{v}_I^2+\mathbf{w}_R^2+\mathbf{w}_I^2=1/2$.
    \item $\mathbf{u}_{R}^2=\mathbf{u}_{I}^2$, $\mathbf{v}_{R}^2=\mathbf{v}_{I}^2$ and $\mathbf{w}_{R}^2=\mathbf{w}_{I}^2$.
    \item $\mathbf{u}_R\cdot\mathbf{v}_R=\mathbf{u}_I\cdot\mathbf{v}_I=\mathbf{w}_R\cdot\mathbf{v}_R=\mathbf{w}_I\cdot\mathbf{v}_I=\mathbf{u}_R\cdot\mathbf{w}_R=\mathbf{u}_I\cdot\mathbf{w}_I=\mathbf{u}_R\cdot\mathbf{v}_I=\mathbf{u}_I\cdot\mathbf{v}_R=\mathbf{w}_R\cdot\mathbf{v}_I=\mathbf{w}_I\cdot\mathbf{v}_R=\mathbf{u}_R\cdot\mathbf{w}_I=\mathbf{u}_I\cdot\mathbf{w}_R=0$.
\end{enumerate}
Naively, 3 contains 12 constraints. However, they are not all independent. 
We consider a generic case where all the vectors are assumed to be non-zero.
We first consider $\mathbf{u}_R$ and $\mathbf{v}_R$ to be independent. As 3 stipulates $\mathbf{u}_R\cdot\mathbf{v}_R =0$, $\mathbf{u}_R$ and $\mathbf{v}_R$ should be perpendicular if they are both non-zero. We next consider $\mathbf{w}_R$, which satisfies $\mathbf{w}_R\cdot \mathbf{u}_R = \mathbf{w}_R\cdot \mathbf{v}_R = 0$. As $\mathbf{w}_R$ is perpendicular to both $\mathbf{u}_R$ and $\mathbf{v}_R$, we have a set of three mutually perpendicular axes in $(\mathbf{u}_R,\mathbf{v}_R,\mathbf{w}_R)$. By the same reasoning, we find that $(\mathbf{u}_R,\mathbf{v}_R,\mathbf{w}_I)$ also forms a set of mutually perpendicular axes. This establishes that $\mathbf{w}_R$ and $\mathbf{w}_I$ are parallel (or anti-parallel) vectors. Following the same line of arguments, we find that $\mathbf{u}_R$, $\mathbf{u}_I$ are a pair of parallel vectors, while $\mathbf{v}_R$, $\mathbf{v}_I$ are also parallel. 

To fix the directions of $(\mathbf{u}_R,\mathbf{u}_I)$, $(\mathbf{v}_R,\mathbf{v}_I)$ and $(\mathbf{w}_R,\mathbf{w}_I)$, we choose a coordinate frame that fixes three mutually perpendicular directions. The choice of frame requires three degrees of freedom, e.g., three Euler angles. 

Below, for concreteness, we fix the direction of $\mathbf{u}$ (including real and imaginary components) to be along $\hat{x}$, that of $\mathbf{v}$ along $\hat{y}$ and $\mathbf{w}$ along $\hat{z}$. We write $\mathbf{u}=\frac{(u+iu')}{2}\hat{x}, \mathbf{v}=\frac{(v+iv')}{2}\hat{y}, \mathbf{w}=\frac{(w+iw')}{2}\hat{z}$. 
From constraint 2, we obtain $u=\pm u', v=\pm v', w=\pm w'$. The constraint 1 yields $u^2+v^2+w^2=1$. With $u=\pm u', v=\pm v', w=\pm w'$, we may write the vectors as $\mathbf{u}=\frac{u}{\sqrt{2}}~e^{\pm i \pi/4} \hat{x}$, $\mathbf{v}=\frac{v}{\sqrt{2}}~e^{\pm i \pi/4} \hat{y}$ and $\mathbf{w}=\frac{w}{\sqrt{2}}~e^{\pm i \pi/4} \hat{z}$. 

We finally arrive at
    \begin{eqnarray}
        \nonumber \mathbf{S}_{m,n} &=&  \frac{u}{\sqrt{2}}~\hat{x} ~\Big\{ e^{i\mathbf{W}_1 \cdot \mathbf{r}_{m,n}\pm i \pi/4} + ~e^{-i\mathbf{W}_1 \cdot\mathbf{r}_{m,n}  \mp i \pi/4} \Big\}+ 
        \frac{v}{\sqrt{2}}~\hat{y} ~\Big\{~e^{i\mathbf{W}_2 \cdot \mathbf{r}_{m,n} \pm i \pi/4} + ~e^{-i\mathbf{W}_2 \cdot \mathbf{r}_{m,n} \mp i \pi/4} \Big\} \\
        &~&+\frac{w}{\sqrt{2}}~\hat{z} ~ \Big\{ ~e^{i\mathbf{W}_3 \cdot \mathbf{r}_{m,n} \pm i \pi/4} + e^{-i\mathbf{W}_3 \cdot \mathbf{r}_{m,n} \mp i \pi/4}\Big\} \nonumber \\
        &=& \sqrt{2}~ u~\cos (\mathbf{W}_1 \cdot \mathbf{r}_{m,n}\pm \pi/4) ~\hat{x}+ 
        \sqrt{2}~v~\cos (\mathbf{W}_2 \cdot \mathbf{r}_{m,n}\pm \pi/4) ~\hat{y}+ 
        \sqrt{2}~w~\cos (\mathbf{W}_3 \cdot \mathbf{r}_{m,n}\pm \pi/4) ~\hat{z},    
\end{eqnarray}
where $u$, $v$ and $w$ are real numbers that satisfy the constraint $u^2 + v^2 + w^2 = 1$.

This ansatz produces a periodic structure with a 16 site motif in one layer, as shown in Fig.~\ref{fig.multiQ}(b) of the main text. From one layer to the next, we fix spins using the tetrahedron rule. It can be checked that this produces the correct ground state energy for any stacking sequence. We count the degrees of freedom in the ground state as follows. To choose a coordinate frame with three mutually perpendicular directions, we require three degrees of freedom (Euler angles). The variables $u$, $v$ and $w$, subject to $u^2+v^2+w^2 =1$, amount to two degrees of freedom. This leads to a five-parameter space of ground states. 

In the discussion above, we have assumed that $\mathbf{u}_R$, $\mathbf{u}_I$, $\mathbf{v}_R$, $\mathbf{v}_I$, $\mathbf{w}_R$ and $\mathbf{w}_I$ are all non-zero. If we allow them to be zero, only one additional class of solutions emerges which can accommodate at most four of the six $\mathbf{W}$ momenta. For a concrete example, we write
\begin{eqnarray}
\nonumber    \mathbf{S}_{m,n} &=& \frac{u}{2} ~(\hat{x} \pm i \hat{y}) ~e^{i\mathbf{W}_1\cdot \mathbf{r}_{m,n}} + \frac{u}{2}~(\hat{x}\mp i\hat{y}) e^{-i\mathbf{W}_1\cdot \mathbf{r}_{m,n}} + \frac{v}{2} ~(1 \pm i )~ \hat{z}~ e^{i\mathbf{W}_2\cdot \mathbf{r}_{m,n}} + \frac{v}{2} ~(1\mp i)~\hat{z} ~e^{-i\mathbf{W}_2\cdot \mathbf{r}_{m,n}}\\
    &=& u \Big\{ \cos(\mathbf{W}_1\cdot \mathbf{r}_{m,n})~\hat{x} ~ \pm  
     \sin(\mathbf{W}_1\cdot \mathbf{r}_{m,n}) ~\hat{y} \Big\} + 
\sqrt{2} ~v~ \cos(\mathbf{W}_2\cdot \mathbf{r}_{m,n} \pm \pi/4) ~\hat{z},
\end{eqnarray}
with the constraint $u^2 + v^2 = 1$. Such solutions form a four-dimensional family of configurations. Three degrees of freedom account for the choice of coordinate frame. We have one more degree of freedom in $u$ and $v$, which are subject to a normalization condition $u^2+ v^2 = 1$. In this class as well, spin orientations on adjacent layers are found using the tetrahedron rule. This yields the same degeneracy on all close-packed lattices.

\section{Degeneracies at phase boundaries}

\begin{figure}[t]
    \centering

    \includegraphics[width=0.329\textwidth]{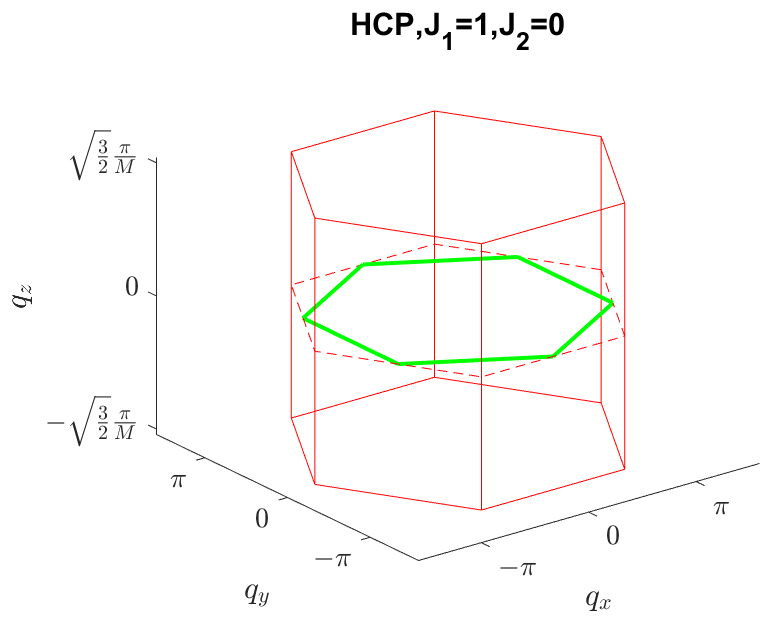}
    \hfill
    \includegraphics[width=0.329\textwidth]{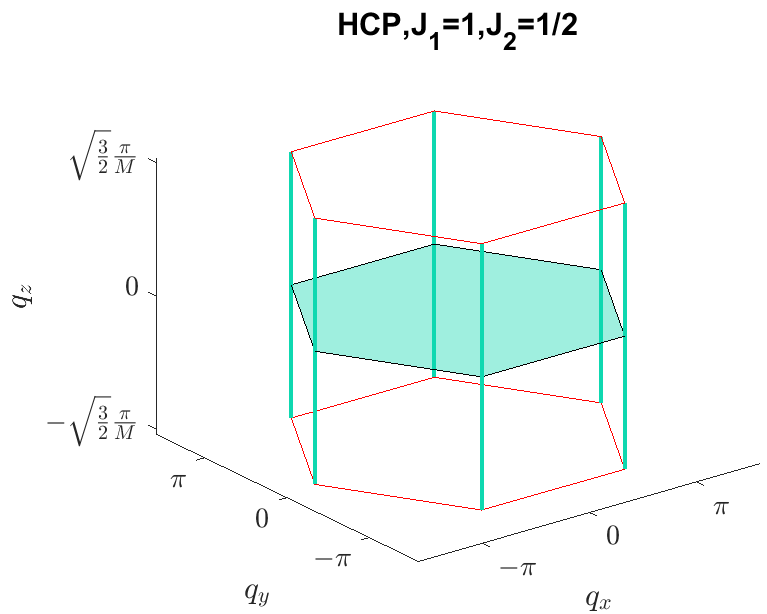}
    \hfill
    \includegraphics[width=0.329\textwidth]{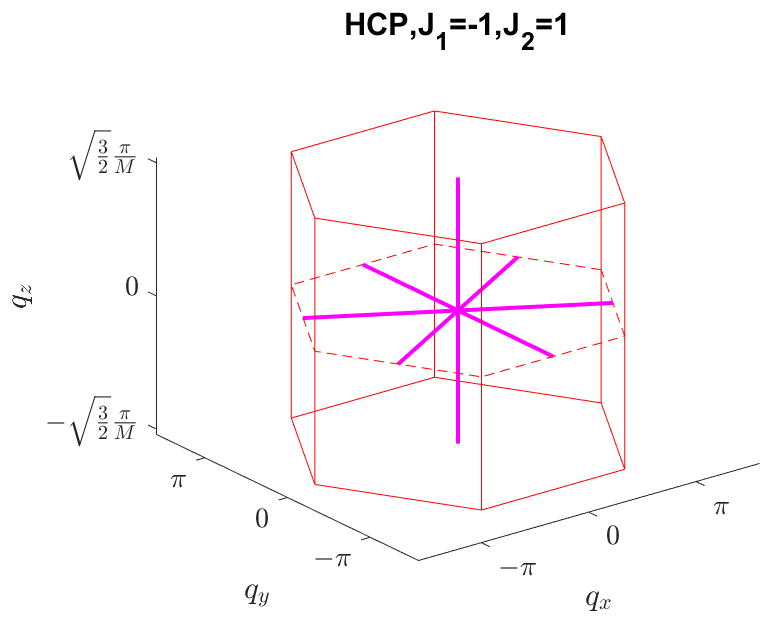}

    \vspace{0.5em}

    \includegraphics[width=0.329\textwidth]{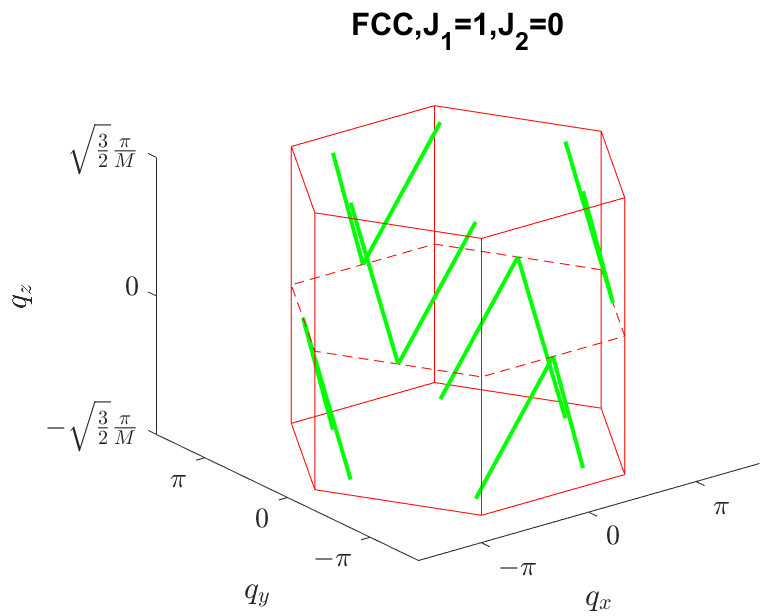}
    \hfill
    \includegraphics[width=0.329\textwidth]{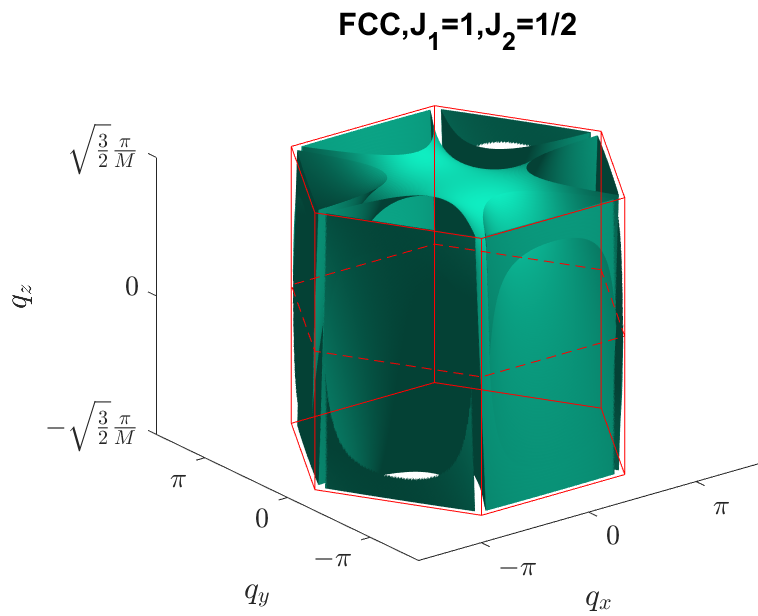}
    \hfill
    \includegraphics[width=0.329\textwidth]{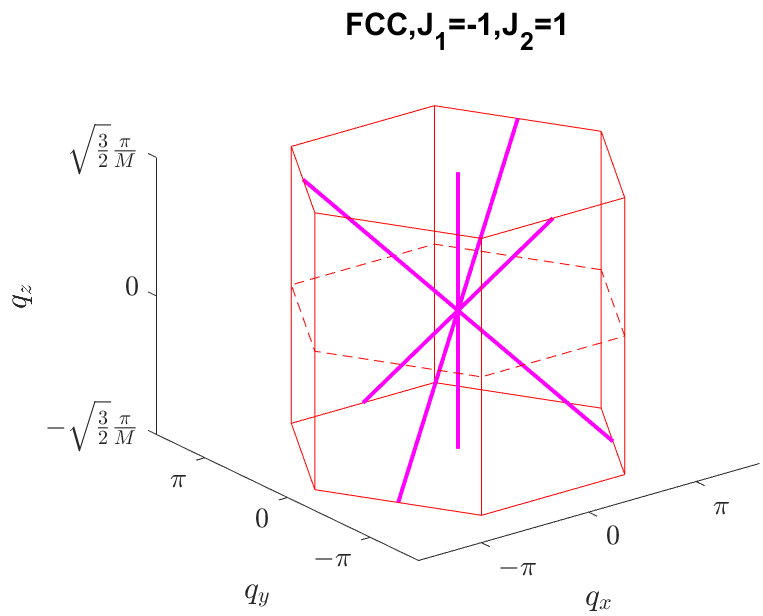}

    \caption{Subextensive LT minimum-manifolds of hcp and fcc lattices at phase boundaries, depicted in the full (unprojected) Brillouin zone. 
    Phase boundaries shown here are $\theta = 0^\circ$ ($J_1=1, J_2=0$), $\theta ~\sim 26.57^\circ$ ($J_1=1, J_2=1/2$) and $\theta=135^\circ$ ($J_1=-1, J_2=1$). At $\theta=0^\circ$ and $135^\circ$, the LT manifold consists of 1D lines, while at $\theta \sim 26.57^\circ$ the manifold consists of both 1D lines and a 2D surface.  
    Here, $M$ is the number of sites in the unit cell and is related to the size of the Brillouin zone in the $q_z$ direction. We have $M=2$ for hcp and $M=3$ for fcc.}
    \label{fig.manifolds}
\end{figure}

The LT minimum-manifolds at three of the four phase boundaries are shown in Fig.~\ref{fig.boundaries} of the main text. They are shown within the projected Brillouin zone, where all close-packed lattices produce the same LT minima. In Fig.~\ref{fig.manifolds}, we show the manifolds for the hcp and fcc lattices in the full, three-dimensional Brillouin zone. The phase boundary at $\theta=270^\circ$ ($J_1=0, J_2=-1$) is not shown as it only has minima at four discrete points within the Brillouin zone. 

\begin{figure}
    \centering
    \includegraphics[width=0.2\linewidth]{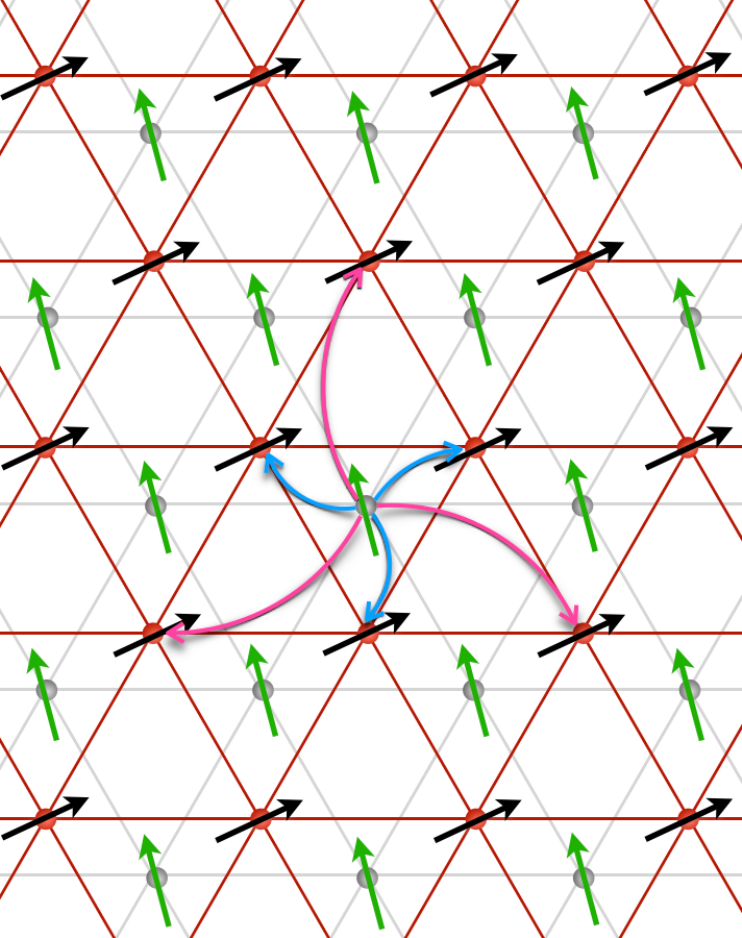}~~~~~~~~~~~~\includegraphics[width=0.2\linewidth]{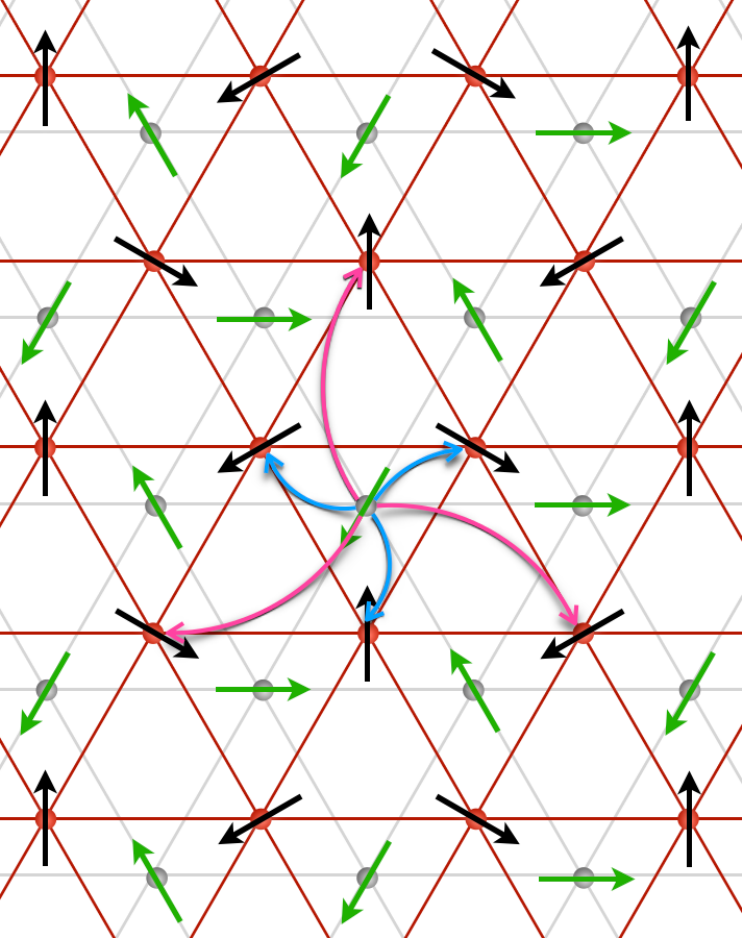}
    \caption{Extended degeneracies at phase boundaries. Left: At $\theta=135^\circ$ ($J_1<0$, $J_2 = -J_1$), each layer can independently order as a ferromagnet. The figure shows two adjacent layers, with the interlayer bonds at the central site marked. The three interlayer $J_1$ bonds and the three interlayer $J_2$ bonds cancel one another as $J_2=-J_1$. This allows each layer to independently pick an ordering direction. Right: At $\theta\sim 26.57^\circ$ ($J_2 = J_1/2 >0$), each layer can independently develop 120$^\circ$ order. The figure shows two layers, with interlayer bonds of the central site marked. The effective field generated at the central site by the interlayer $J_1$ couplings vanishes, as does that of interlayer $J_2$ couplings. }
    \label{fig.extended}
\end{figure}

The extended ground state degeneracy at $\theta=0^\circ$ is discussed in the main text -- see Fig.~\ref{fig.theta0}. Here, we discuss the extended degeneracies that appear at $\theta=135^\circ$ and at $\theta\sim 26.57^\circ$. At $\theta=135^\circ$, we have $J_1<0$ and $J_2>0$ with $J_1$ and $J_2$ having the same magnitude. The ground state energy at this point is given by $-3J_1$ per spin, with the spin length set to unity. We focus on the line of minima that is parallel to the stacking axis. This line of minima signifies that each triangular-lattice layer can order independently as a ferromagnet (in any close-packed structure). With six $J_1$ neighbours within each layer, the intra-layer couplings account for the ground state energy completely. The out-of-plane couplings effectively cancel out, as shown in Fig.~\ref{fig.extended}(left). Each spin couples to six neighbours in an adjacent layer. Three of these are $J_1$ neighbours while three are $J_2$-neighbours. All six neighbours have spins with the same orientation. As $J_2 = -J_1$, the interlayer $J_1$- and $J_2$-energy contributions cancel out.

At $\theta\sim 26.57^\circ$, we have $J_2 = J_1/2$ with both $J_1$ and $J_2$ being positive. The LT ground state energy comes out to be $-3J_1/2$ per site. At this point, there are two families of LT minima. The first is a pair of straight lines that are parallel to the stacking axis, passing through the corners ($K$ points) of the projected Brillouin zone. These lines can be interpreted as independent $120^\circ$ ordering in each layer (in any stacking sequence). The ground state energy is fully accounted for by the intra-layer couplings. The interlayer couplings effectively vanish in any close-packed structure as shown in Fig.~\ref{fig.extended}(right). 

\begin{figure}
    \centering
    \includegraphics[width=0.3\linewidth]{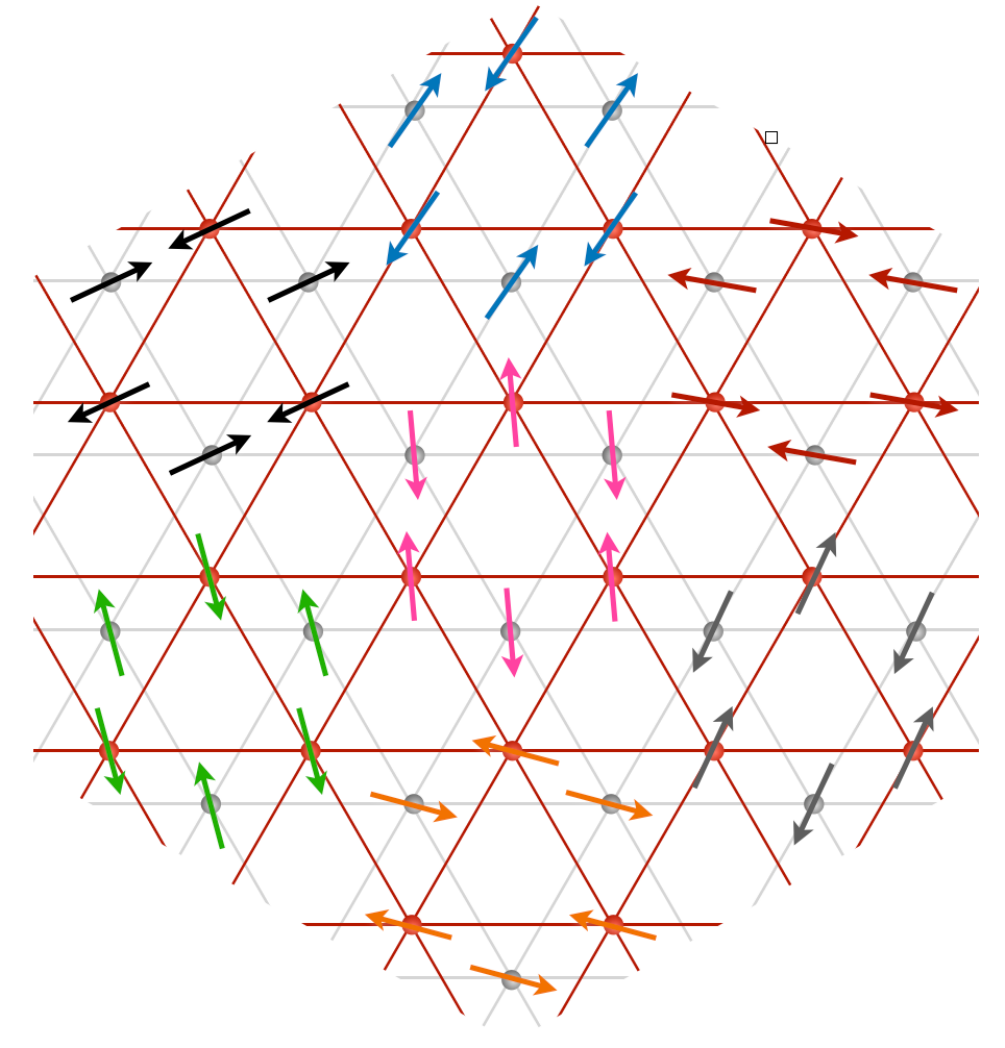}
    \caption{Extensive degeneracy in the hcp lattice at $\theta\sim 26.57^\circ$ ($J_2 = J_1/2 > 0$). Seen along the stacking axis, the hcp lattice can be viewed as a collection of pillars formed by face sharing octahedra. On each pillar, we may assign an independent N\'eel ordering vector. A and B layers are shown with spin orientations on the top sites. Below each visible site, we have a line of sites that all carry the same spin orientation. All such configurations are classical ground states. }
    \label{fig.pillars}
\end{figure}

The second family is a surface that stretches over the entire projected Brillouin zone. In the three dimensional Brillouin zone of the hcp lattice, this family corresponds to a flat band of minima in the $q_z =0$ plane -- see Fig.~\ref{fig.manifolds}. In the fcc lattice, however, it corresponds to a complex surface as discussed in Ref.~\cite{Balla2020} within the conventional fcc Brillouin zone. On all close-packed lattices, this surface can be understood as describing extensive degeneracy that arises from a local constraint. Any close-packed lattice can be viewed as network of face-sharing octahedra. In this point of view, each $J_1$ bond forms an edge of two adjacent octahedra. However, each $J_2$ bond is contained within a single octahedron. With $J_2 = J_1/2$, the Hamiltonian can be written as 
\begin{eqnarray}
    H = \sum_{octa.} \vert \mathbf{S}_{octa.}^{tot.}\vert^2 + \mathrm{constants},
\end{eqnarray}
where the sum runs over all octahedra. The vector $\mathbf{S}_{octa.}^{tot.}$ represents the sum of spins on the octahedron. With this form, the ground state is achieved by ensuring that $\mathbf{S}_{octa.}^{tot.} =0$ on all octahedra. This can evidently lead to large ground state degeneracy. A particularly exciting example can be seen in the hcp lattice, with (AB) stacking. We may divide this lattice into parallel one-dimensional `pillars'. Each pillar can then develop N\'eel order independently as shown in Fig.~\ref{fig.pillars}, with every octahedron having vanishing total spin. The independent pillars can be viewed as local degrees of freedom that arise from the flat band of LT minima.

\section{Order by disorder}

The gauge symmetry discussed in the main text connects classical ground states across close-packed structures. 
Unlike the ground state itself, there is no symmetry that relates spin wave fluctuations in one structure with those of another. To demonstrate this, we carry out standard Holstein Primakov spin wave analysis which expands the Hamiltonian in powers of the spin length, $S$. The dominant $\mathcal{O}(S^2)$ term is the classical ground state energy, obtained from LT analysis. The leading correction, at $\mathcal{O}(S)$, is the zero-point energy contribution of spin waves\cite{auerbach2012interacting} . 

\begin{figure}
    \centering
    \includegraphics[width=0.8\linewidth]{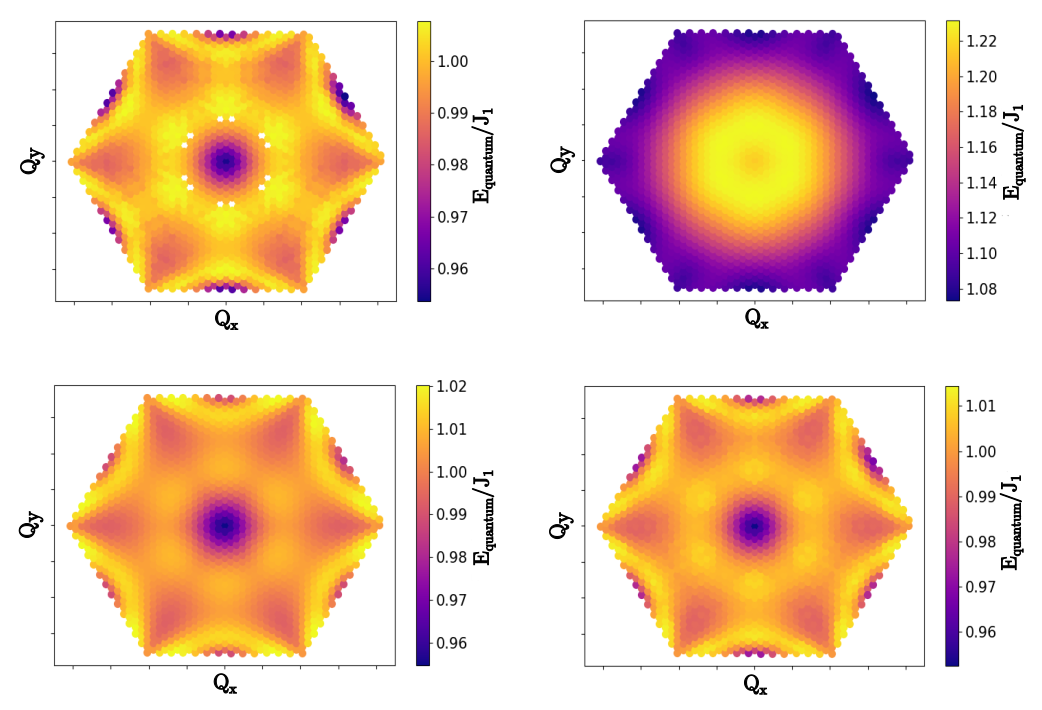}
    \caption{Zero-point energies over the surface of classical ground states at the $\theta\sim 26.57^\circ$ point. Top left is for fcc, top right is for hcp, bottom left is for dhcp and bottom right is for an eight-layer stack (ABCBCABC). The spin wave energies are clearly different from one structure to the next. Crucially, the minimum of the $\mathcal{O}(S)$ term appears at different locations. In the fcc lattice, minima occur at the $\Gamma$ and $M$ points. In the hcp case, minima appear solely at the $M$ points. For dhcp and the eight-layer stack, there is a unique minimum at the $\Gamma$ point.}
    \label{fig.ObD}
\end{figure}

We plot the $\mathcal{O}(S)$ zero-point-energy correction in Fig.~\ref{fig.ObD}. For concreteness, we focus on the $\theta\sim 26.57^\circ$ point in the phase diagram which yields a surface of LT minima. The three-dimensional surface may differ from one close-packed structure to another. However, its projection onto the $q_a-q_b$ plane is the same in every structure. All points on this surface have the same energy at $\mathcal{O}(S^2)$, representing the classical energy contribution. The figure plots the $\mathcal{O}(S)$ correction over the surface of LT minima, projected onto the stacking plane. The $\mathcal{O}(S)$ energy is manifestly different from one structure to the next. In particular, the $\mathcal{O}(S)$ correction has minima that differ from one structure to the next. That is, the $\mathcal{O}(S)$ correction `selects'  different states in different structures.

\begin{figure}
    \centering
    \includegraphics[width=0.8\linewidth]{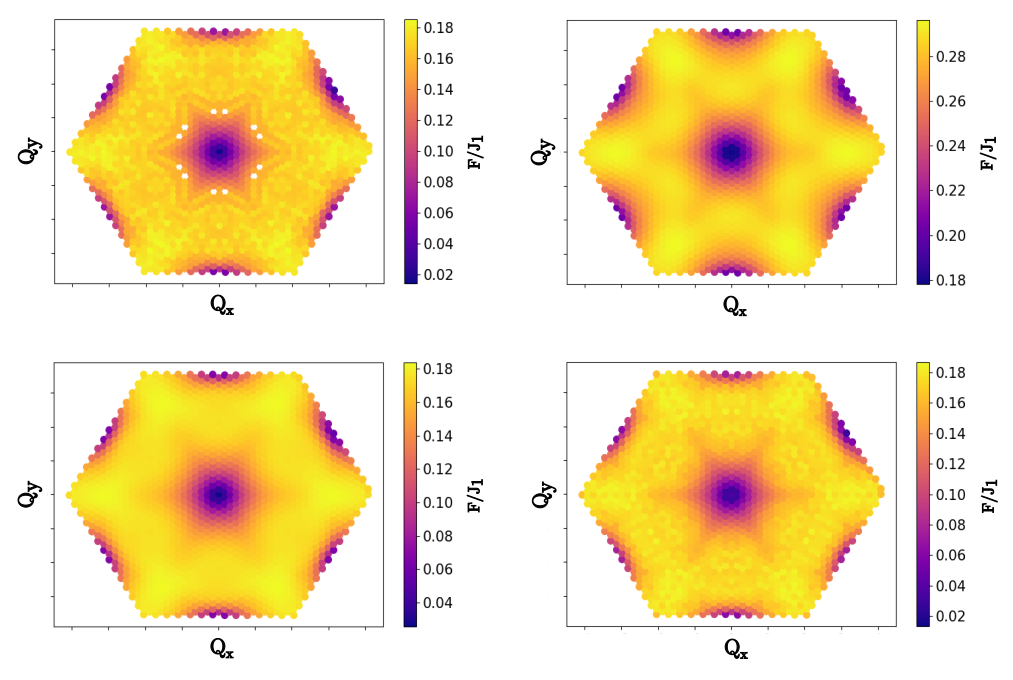}
    \caption{The free energy of spin wave fluctuations over the surface of classical ground states at the $\theta\sim 26.57^\circ$ point. Top left is for fcc, top right is for hcp, bottom left is for dhcp and bottom right is for an eight-layer stack (ABCBCABC). As with the quantum order-by-disorder case, the minimum of the free energy appears at different locations. In the fcc, the minima are at the $\Gamma$ and $M$ points. For hcp, dhcp and the eight-layer stack, there is a unique minimum at the $\Gamma$ point.}
    \label{fig.tObD}
\end{figure}

As is well known, order by disorder selection may also occur due to thermal fluctuations. This is typically described by calculating the free energy due to spin wave fluctuations about each classical ground state. In Fig.~\ref{fig.tObD}, the free energy (in the limit of low temperatures) is plotted for three different close-packed structures. The magnitudes differ, illustrating that there is no symmetry connecting thermal fluctuations. Furthermore, the `selected' state or the minimum of the free energy changes from one structure to another.

\end{document}